\documentclass[aps,prd,a4paper,twocolumn,showpacs,showkeys,fleqn,floatfix]{revtex4-1}

\usepackage{graphicx}
\usepackage{color}
\usepackage{amsmath}

\usepackage[cp1251]{inputenc}

\newcommand{\figwidth}{0.98\columnwidth}

\newcommand{\vect}[1]{\mathbf{#1}}
\newcommand{\phcc}{(C$_4$H$_{12}$N$_2$)Cu$_2$Cl$_6$}
\newcommand{\phbc}{(C$_4$H$_{12}$N$_2$)Cu$_2$Br$_{6y}$Cl$_{6(1-y)}$}
\newcommand{\boldeta}{\boldsymbol\eta}
\newcommand{\dimpy}{(C$_7$H$_{10}$N$_2$)$_2$CuBr$_4$}
\newcommand{\dimpyzn}{(C$_7$H$_{10}$N$_2$)$_2$Cu$_{1-x}$Zn$_x$Br$_4$}
\newcommand{\sul}{Cu$_2$Cl$_4 \cdot$H$_8$C$_4$SO$_2$}
\begin{document}

\title{Magnetic resonance of collective paramagnets with gapped excitations spectrum.}

\author{V.N.Glazkov}
\email{glazkov@kapitza.ras.ru}
\affiliation{P. L. Kapitza Institute for Physical Problems RAS, ul. Kosygina  2, 119334 Moscow, Russia}

\begin{abstract}
\emph{This text is an expanded author-translated version of the review prepared for the memorial issue of Journal of Experimental and Theoretical Physics dedicated to 100-year anniversary of A.S.~Borovik-Romanov (full memorial volume is available as JETP \textbf{131} No.1 (2020) [ZhETF \textbf{158} No.1 (2020)])}\\ \\Some magnets due to particular geometry of the exchange bonds do not undergo transition to the conventional magnetically ordered state despite of the presence of significant exchange couplings. Instead, a collective paramagnetic state is formed. The later state can remain stable down to $T=0$ if the ground state of this magnet turns out to be nonmagnetic singlet separated from the excited triplet states by an energy gap. Low-temperature spin dynamics of the collective paramagnets with gapped excitations spectrum (or spin-gap magnets) can be described in terms of a dilute gas of the triplet excitations. Applied magnetic field can suppress the energy gap, resulting in the formation of the gapless spin-liquid state or even leading to the unusual phenomenon of field-induced antiferromagnetic order. Introduction of defects in the crystallographic structure of the spin-gap magnet can result either in the formation of multi-spin paramagnetic center or in the formation of randomly distributed modified exchange bonds in the crystal. This review includes results of electron spin resonance (ESR) spectroscopy study of several representative quantum paramagnets  with gapped excitations spectrum: quasy-two-dimensional magnet \phcc{}, quasy-one-dimensional magnets of ``spin-tube'' type \sul{} and ``spin-ladder'' type \dimpy{}. We will demonstrate that ESR absorption spectra reveal common features of these systems:  ESR spectroscopy allows to observe and characterize fine structure if the triplet energy levels, to identify many-particles relaxation processes in the gas of triplet excitations and to observe collective spin-wave oscillations in the field induced antiferromagnetically ordered state, as well as to observe some individual features of the studied systems.
\end{abstract}

\date{\today}

\maketitle
\tableofcontents

\section{Introduction}

Development of electron spin resonance technique have started in mid-40-th from the discovery of magnetic resonance in paramagnets by E.K.~Zavoisky \cite{zavoiskii}. Conventional paramagnetic resonance is observed when the resonance absorption is due to the induced transitions between the Zeeman split spin sublevels of individual ions an interaction of magnetic ions is weak. Even being applied to simple paramagnets, electron spin resonance is a sensitive spectroscopic technique with unique abilities: difference of $g$-factor of different magnetic ions and anisotropy of $g$-tensor in crystal allows to resolve signals from different ions or from the ions in different crystalline surroundings, the shape of the resonance absorption spectra and the width of the absorption line give insight into spin-spin interactions or to the interaction of magnetic ion with crystal field \cite{altkoz,ablin}.

Application of ESR spectroscopy to the antiferromagnetically ordered crystals have developed since 1950-ies \cite{cucl2-first,kittel,keffler,NYK}. A.S.~Borovik-Romanov was one of the pioneers of these studies, he has investigated magnetic resonance in weak ferromagnets, nonlinear effects in magnetically ordered crystals, parametric excitation of the spin-waves in antiferromagnets and coupled electron-nuclear oscillation modes  \cite{br1,br2,br3,br4,br5,br6,br7,br8,br9}. Contrary to the case of simple paramagnets, properties of the ordered antiferromagnets are governed by opposite hierarchy of interactions: the strongest interaction is the Heisenberg exchange coupling resulting in the formation of the regular pattern of ordered spins. Microwave absorption by ordered antiferromagnet in this case can not be interpreted  as a flip of individual spin (transition between spin sublevels of the individual ion) --- instead, microwave field excite collective excitations of antiferromagnetic spin structure (spin waves). Since ESR experiments are usually carried out at centimeter or millimeter wavelength (typical microwave frequencies 10--100~GHz), photon wavelength far exceeds the interatomic distances and absorption of microwave radiation excites spin waves with $q\simeq 0$. High energy resolution of the ESR spectroscopy (routine resolution is less than 1~GHz) makes it a very informative method to study low-energy spin dynamics, to determine  gaps in the magnon spectrum related to anisotropy, to detect phase transitions accompanied by the change of spin structure \cite{turov,gurevich}. Antiferromagnet eigenfrequencies and the quantity of eigenmodes is related to the type of magnetic ordering \cite{andmar}.

Recently one of the focus topics of the research in the field of magnetism have shifted to the systems, which despite of the strong antiferromagnetic exchange coupling do not demonstrate conventional Neel ordering down to $T\ll \Theta$ (here $\Theta$ is a Curie-Weiss temperature) due to particular geometry of the exchange bonds. Low temperature state of these systems is a collective paramagnetic state. One of the simplest examples of such a system is that of  the weakly coupled antiferromagnetic dimers (pairs of $S=1/2$ spins with antiferromagnetic couplings). Ground state of the isolated dimer is a singlet ($S=0$), and energy of an excited triplet state ($S=1$) is larger by the value of exchange coupling constant $J$. Weak inter-dimer coupling does not change the ground state, but excited triplet states becomes delocalized and acquires dispersion. There are examples of other mechanisms of energy gap formation in a collective paramagnet, e.g., in a one-dimensional ``spin-ladders'' or in a one-dimensional chains of integer spins (Haldane magnets). Common feature of these systems is their stability against formation of the  conventional antiferromagnetic order if there is a finite gap  $\Delta$ between the ground singlet state and excited triplets  \cite{kolezhuk,zapf,smirglazjetp,schmidiger-phd}. At low temperatures $T\ll\Delta$, population numbers of the triplet states are low and magnetic properties of a spin-gap magnet can be described as the properties of a dilute gas of almost free quasiparticles --- triplons. Triplons have a spin $S=1$, they determine all low temperature magnetic properties of the spin-gap magnet, including the resonance absorption in ESR experiments. Thus, in some sense, these systems ``complete the circle'' of the development of electron spin resonance, describing the properties of a   \emph{collective} paramagnet in the same terms as the traditional paramagnetic resonance of isolated   \emph{local} spins.

As temperature changes a series of crossovers takes place, first from the regime of the dilute gas of triplons ($T\ll\Delta$) to the regime of collective oscillations of strongly coupled spins ($T\sim\Theta$) and than to high-temperature paramagnetic limit ($T\gg\Theta$). Description of spin-dynamics over the complete temperature range (including the spin relaxation rate) could be, in principle, described by a compact set of microscopic parameters. However, there is no general theory applicable over the full temperature range, and description of certain model systems is known in some of the limiting cases \cite{oshiaff,furuya-jpsj}.

Another problem of interest is the stability of this collective paramagnetic state under various external perturbations. We will limit our consideration here to the effects of external magnetic field and controlled introduction of the impurities only. Under applied magnetic field triplet sublevels split in the whole Brillouine zone and at a certain field $H_{c1}\simeq \Delta/(g\mu_B)$ energy of the lower triplet sublevel equals the singlet state energy. A quantum phase transition to the gapless state, and sometimes even to the antiferromagnetically ordered state, takes place at this field \cite{zapf}. Magnetization of spin-gap magnet saturates at a larger field  $H_{c2}\simeq\Theta/(g\mu_B)$. Nonmagnetic impurities can occupy positions either of the magnetic ion, thus depleting the magnetic subsystem, or of the nonmagnetic ion mediating formation of superexchange bonds. Substitution of one of the ions by an impurity results either in the formation of new paramagnetic center formed by several spins of spin-gap magnet matrix (see, e.g., \cite{smirglazjetp}), or in the appearance of exchange bonds with varying  strength randomly distributed through the lattice \cite{glassy,bg-exp,random1,random2,random-exp}.

One of the first applications of ESR spectroscopy to the study of spin-gap magnets was the study of quasi-one-dimensional Haldane magnets  NENP and NINO  by M.~Date and K.~Kindo \cite{date}.

We will discuss in this review the results of recent studies of several typical spin-gap magnets: quasy-two-dimensional dimer magnet PHCC (\phcc{}) demonstrating ESR absorption spectra, which can be fully described at $H<H_{c1}$ in the model of free triplons in an anisotropic crystal; its bromine-diluted isostructural compound \phbc{}, which is an example of formation of $S=1$ paramagnetic centers on nonmagnetic dilution of spin-gap magnet; quasi-one-dimensional magnet with the geometry of exchange bonds of ``spin-tube'' type sul-Cu$_2$Cl$_4$ (\sul{}), demonstrating incommensurate position of the spectrum minimum at $H<H_{c1}$ and a helical magnetic ordering at $H>H_{c1}$; and a ``spin-ladder'' magnet DIMPY (\dimpy{}), which is a rare example of a strong-leg spin ladder, along with its partially depleted analogue \dimpyzn{}, which, unexpectedly, demonstrates narrowing of the ESR line on doping.

\section{Magnetic resonance of a spin-gap magnet: below $H_{c1}$ and above $H_{c1}$.}

\begin{figure}
  \centering
  \includegraphics[width=\figwidth]{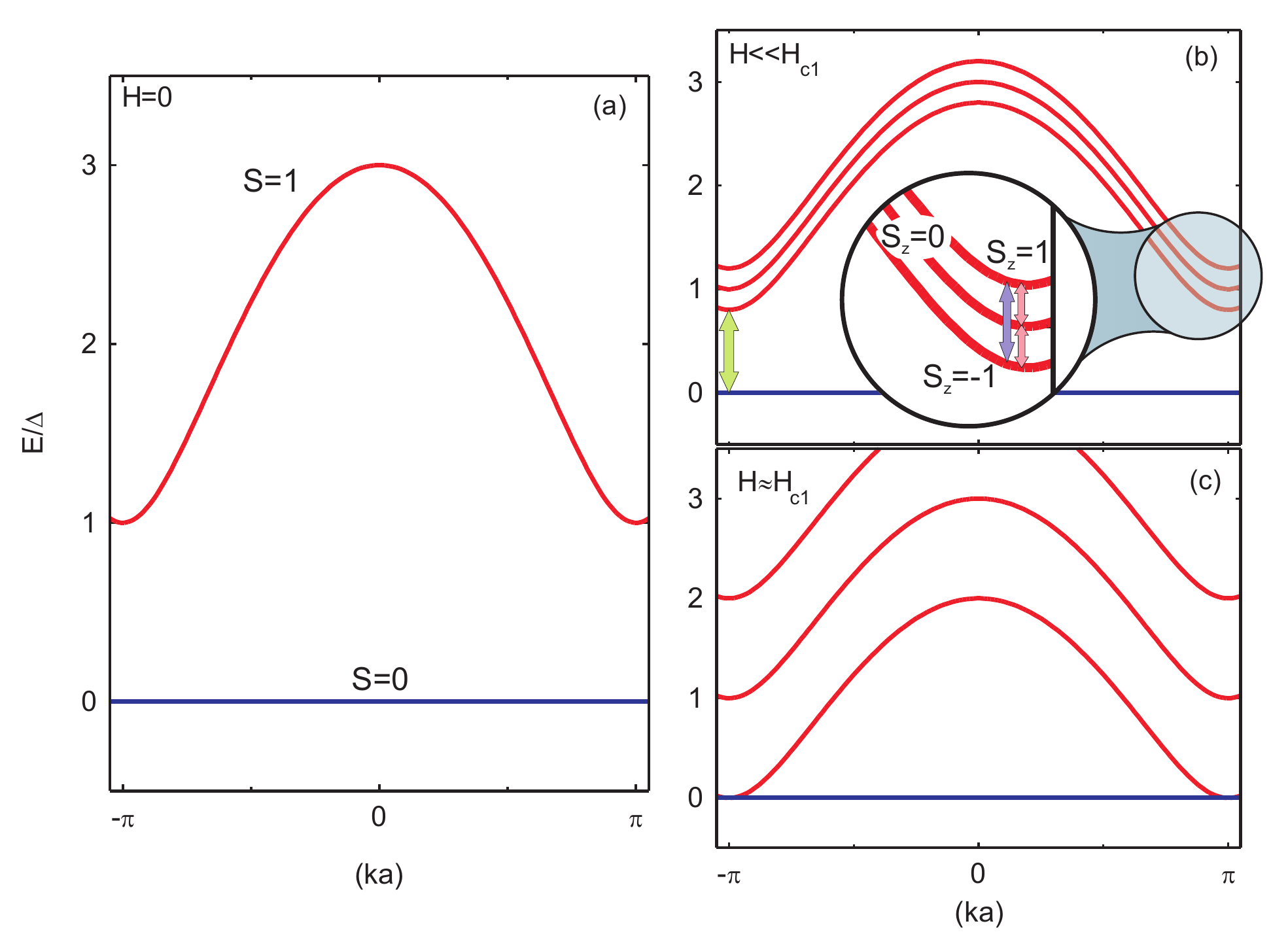}
  \caption{Scheme of the elementary excitations of the spin-gap magnet neglecting anisotropic spin-spin interactions. (a) At zero magnetic field (b) At  field $H\ll H_{c1}$ (small field limit, range of perturbative approach applicability). Arrows indicate possible transitions between spin sublevels with $\Delta q=0$, $\Delta S_z=1$ and   $\Delta S_z=2$, as well as singlet-triplet transition at $k a=\pi$, which can be allowed in the presence of staggered field. (c) Qualitative behavior close to $H_{c1}$.}\label{fig:spectra-scheme}
\end{figure}

Let us first discuss specific features of electron spin resonance in a spin-gap system. We assume that there is a gap $\Delta$ in the excitation spectrum which separates singlet ground state from the excited triplets. Typically, minimum of the triplet excitations energy  $E(q)$ is found at the antiferromagnetic wavevector (Fig.\ref{fig:spectra-scheme}-a). At low temperatures $T\ll\Delta$ magnetic susceptibility of such a magnet is proportional to the number of thermally activated triplons:

\begin{equation}\label{eqn:chi(T)}
    \chi(T) \propto \int \exp\left(-\frac{\Delta+\alpha q^2}{T}\right) d^{D} q \propto T^{D/2} e^{-\Delta/T},
\end{equation}
\noindent here $D$ is the spacial dimensionality of the spin system.

Integral intensity of ESR absorption is proportional to the static magnetization. Microwave field is practically uniform in ESR experiment since linewidth $\lambda\sim$ 1 cm and a photon wavevector $k_{ph}\ll \pi/a$. However, this does not forbid to observe resonance absorption due to transitions between spin sublevels of triplet excitations close to  minimum of the excitations spectrum $E(q)$ (located away from the Brillouine zone center and usually close to the antiferromagnetic wavevector) since  $\Delta q=0$ for the transitions between the states with the same momentum (Fig. \ref{fig:spectra-scheme}-b).

Within the exchange approximation (neglecting the anisotropic spin-spin couplings) triplet excitations are degenerated over the spin projection at zero field. In a real crystals joint effects of spin-orbital coupling and crystal field (originating from gradients of electric field at  magnetic ion position) produce anisotropic spin-spin interactions. Traditionally these interactions are expressed as a sum of symmetric anisotropic interaction $\sum_{\alpha\beta} G_{\alpha\beta} {\hat S}_{1\alpha}{\hat S}_{2\beta}$ (here $G_{\alpha\beta}=G_{\beta\alpha}$ usually constrained as  $Tr G=0$)and antisymmetric Dzyaloshinskii-Moriya interaction $\vect{D}\left[{\hat{\vect{S}}}_1\times{\hat{\vect{S}}}_2 \right]$. These anisotropic interactions reduce symmetry of the spin Hamiltonian and lift degeneracy over the spin projection resulting in the zero-field splitting of triplet sublevels. This effect was observed firstly for quasi-one-dimensional Haldane magnets NENP and NINO by M.~Date and K.~Kindo \cite{date}.

At low fields effect of anisotropic spin-spin interactions can be treated perturbatively yielding an effective anisotropy acting on an $S=1$ triplet quasiparticle. Taking into account $g$-factor anisotropy, effective Hamiltonian of a triplon can be written as:

\begin{equation}\label{eqn:aniseff}
{\hat{\cal H}}_{eff}=\Delta(\vect{k})+\vect{H}\cdot \tilde{g} \cdot {\hat{\vect{S}}}+D_{\vect{k}} {\hat S}_Z^2+E_{\vect{k}}\left({\hat S}_X^2-{\hat S}_Y^2 \right),
\end{equation}

\noindent here effective anisotropy parameters  $D_{\vect{k}}$, $E_{\vect{k}}$ and orientation of the $XYZ$ basis depend on exact details of anisotropic spin-spin interactions and on the symmetry of the crystal. Effective anisotropy parameters  $D_{\vect{k}}$ and $E_{\vect{k}}$ could depend also on triplon wavevector $\vect{k}$. Zero-field energies of the triplet sublevels are then $\Delta(\vect{k})$ and $\Delta(\vect{k})+D_{\vect{k}}\pm E_{\vect{k}}$. Within framework of perturbative approach problem of triplet sublevels field dependence is fully equivalent to the classical problem of spin $S=1$ in a crystal field \cite{altkoz}. Zero-field splitting results in the splitting of ESR absorption line in two components. Anisotropic spin-spin interactions also weakly allow ``two-quantum'' transition $\Delta S_z=2$, which has a resonance field approximately equal to half of the conventional magnetic resonance field. As a result, fine structure of ESR absorption spectra including several spectral components appears.

Besides of zero-field splitting of triplet sublevels anisotropic spin-spin interactions could allow ESR transitions between the  $S=0$ and $S=1$ states. Different spacial parity of these states requires presence of the interaction antisymmetric on spin permutation (i.e., Dzyaloshinskii-Moriya interaction) to mix these states. Additionally, spacial uniformity of the singlet ground state  ($q_{S=0}=0$) means that singlet-triplet transition will take place to  $q=0$ excited state --- which is away from the spectrum minimum. Singlet-triplet transitions to  $\vect{q}\neq 0$ triplet states are possible only if anisotropic spin-spin interactions are spatially modulated with the same wavevector $\vect{q}$. For example, staggered Dzyaloshinskii-Moriya interaction could allow transitions at a wavevector $q=\pi/a$ of a one dimensional magnet (Fig. \ref{fig:spectra-scheme}-c). Observation of such singlet-triplet transitions allow to follow directly field dependence of triplet sublevels up to the critical field $H_{c1}$.

Two general approaches have been developed to describe field dependence of triplet sublevels. Fermionic approach developed by Tsvelik \cite{tsvelik} describes exactly one-dimensional systems which do not order at $H>H_{c1}$. Formally, results of this approach coincide with the results of perturbative treatment up to the critical field \cite{zaliznyak}. Bosonic or macroscopic treatment \cite{Affleck0,Affleck,farmar} can be applied to the systems which order above the critical field at sufficiently low temperatures. Parameters of both models can be tuned to yield the same energy levels at low fields, however these models predict different effects of anisotropic spin-spin interactions on the energy of triplet sublevels close to the critical field $H_{c1}$: within fermionic approach energy of triplet sublevels is linear with field up to complete closing of the energy gap, while bosonic model predicts nonlinear field dependence of the triplet sublevels as the magnetic field approaches critical value.

After the gap is closed at $H_{c1}$ gapless (in the exchange approximation) state remains up to the saturation field $H_{c2}$. If exchange bonds form a 3D pattern (i.e., if there is a weak coupling between the low-dimensional spin subsystems), field induced antiferromagnetism appears at  $H_{c1}<H<H_{c2}$ at sufficiently low temperature. This field induced transition to the ordered state was discussed in connection with formal similarity with the phenomenon of Bose-Einstein condensation \cite{zapf,giamarchi-nature}. This phase transition was observed at various spin-gap systems, it can be observed as a typical $\lambda$-anomaly on $C(T,H)$ curves or as an appearance of a magnetic Bragg peak (see, e.g., Refs. \onlinecite{tlcucl3-1,tlcucl3-2} for prototypical dimer magnet TlCuCl$_3$). Spin-waves excitations of the ordered phase at $H_{c1}<H<H_{c2}$ could give rise to a specific mode of antiferromagnetic resonance \cite{glazkov-tlcucl3,glazkov2-tlcucl3}. Compared to the usual antiferromagnetic resonance in a two-sublattices antiferromagnet \cite{gurevich}, one could note that at sufficiently high fields $H>H_{c1}\gg \sqrt{H_A H_E}$, here $H_A$ and $H_E$ are the field of anisotropy and exchange field, frequency of one of the resonance modes is close to the Larmor frequency  $\omega\simeq\gamma H \gg \gamma\sqrt{H_A H_E}$, while the second, low-frequency, resonance mode $\omega \sim \gamma\sqrt{H_A H_E}$ corresponds to the field-independent resonance mode of a conventional antiferromagnet. Within the traditional mean field model frequency of this mode is also proportional to the order parameter amplitude, which is field-independent in ordinary antiferromagnet at $H\ll H_{sat}$. In the case of field-induced ordering order parameter develops from zero at $H_{c1}$, which makes this mode field-dependent. Since close to the critical field order parameter is far from saturation, longitudinal oscillations of order parameter are also possible. Bosonic (macroscopic) model can describe this eigenmode above $H_{c1}$ \cite{farmar}.

Simple picture described above is valid at low temperatures $T\ll\Delta$, when population numbers of triplet excitations are low. As temperature increases excitations population numbers increase as well and triplon-triplon repulsion has to be taken into account (dominating antiferromagnetic excitations do not favor existence of two triplons on adjacent lattice sites), which results in spectrum renormalization and in the temperature dependence of the critical field. This complicates the problem and compact description of spin-gap magnet at $T\sim\Delta$ is not possible. On further heating at $\Delta<T<\Theta$ (here $\Theta$ is the Curie-Weiss temperature, corresponding to the characteristic scale of the exchange interaction), spin dynamics corresponds to strongly correlated but not ordered magnet, and finally at  $T\gg\Theta$ spin-gap magnet enters the usual high-temperature paramagnetic limit.

ESR linewidth is determined by relaxation time of spin excitations. At high temperatures $T\gg\Theta$ ESR linewidth  can be described within the framework of conventional van Vleck theory  (see, e.g., \cite{altkoz}). Angular dependencies of ESR linewidth allow to identify anisotropic spin-spin interactions responsible for spin relaxation, to determine their magnitudes and orientation of their main axes. Note, that in principle the same anisotropic interactions are responsible for the zero-field splitting of triplet sublevels discussed above as well as for the possible singlet-triplet ESR transitions. 

On cooling spin relaxation became affected by formation of short-range correlations. Corresponding corrections to the ESR linewidth can be found as high temperature expansion series \cite{soos,faizulin}. Alternatively,  M.~Oshikawa and J.~Affleck have developed theory \cite{oshiaff} applicable to low-dimensional magnets at $T \lesssim \Theta$. An important qualitative result of Refs. \onlinecite{oshiaff,faizulin} is that for symmetric anisotropic spin-spin interaction linewidth decreases on cooling, while for antisymmetric Dzyaloshinskii-Moriya interaction linewidth should increase on cooling. Temperature dependence of ESR linewidth in spin ladder was considered in Ref. \onlinecite{furuya-jpsj} for the case of symmetric anisotropic interaction both on the legs and on the rungs of the spin ladder. It turned out that temperature dependence of the ESR linewidth follow predictions of  \cite{oshiaff} (linewidth decrease on cooling) when only  symmetric anisotropic interaction along the legs is taken into account,  while the symmetric anisotropic interaction on the rungs yields contribution increasing on cooling  $\propto (1/T)$ and reaching maximum of linewidth at  $T\simeq \Delta$. 

Finally, at  $T<\Delta$ spin-gap magnet enters particular spin-relaxation regime: concentration of triplons freezes out and limit of ideal dilute gas of quasiparticles is reached. One expect that triplon lifetime increases on cooling and ESR linewidth decreases. If ESR linewidth is limited by certain triplon-triplon interaction, probability of this process should be determined by their concentration, i.e. quasiparticles lifetime should be proportional to $\rho^{-(n-1)}$, here  $\rho\propto e^{-\Delta/T}$ is the triplons concentration and integer $n$ is the number of triplons that have to interact simultaneously to contribute effectively to spin relaxation. 

Additional maximum on ESR linewidth temperature dependence can appear if fine structure od ESR line became resolved at some temperature. This maximum is related to the transition from the regime of exchange-narrowed ESR spectrum with unresolved fine structure to regime of ESR spectrum with resolved fine structure \cite{chestnut}.  On approaching this maximum from high-temperatures (from the regime of exchange-narrowed line) linewidth $\Delta H \simeq \delta^2/(J_{eff}/(g\mu_B))$, here $\delta$ --- splitting of ESR spectrum fine structure and  $J_{eff}$ --- effective exchange coupling constant which is roughly proportional to triplons concentration.

One should also note that high-frequency \cite{ozeroff} or high-field \cite{krasnikova} ESR-active transitions of particular nature can exist in some spin-gap magnets, but we will not consider these transitions in details in this review.

\section{Application of ESR spectroscopy to the study of collective paramagnets with gapped excitations spectrum}
\subsection{Quasy-two-dimensional magnet \phcc{}: fine structure of triplet sublevels and field-induced antiferromagnetic ordering}

\begin{figure}
\centering
\includegraphics[width=\figwidth]{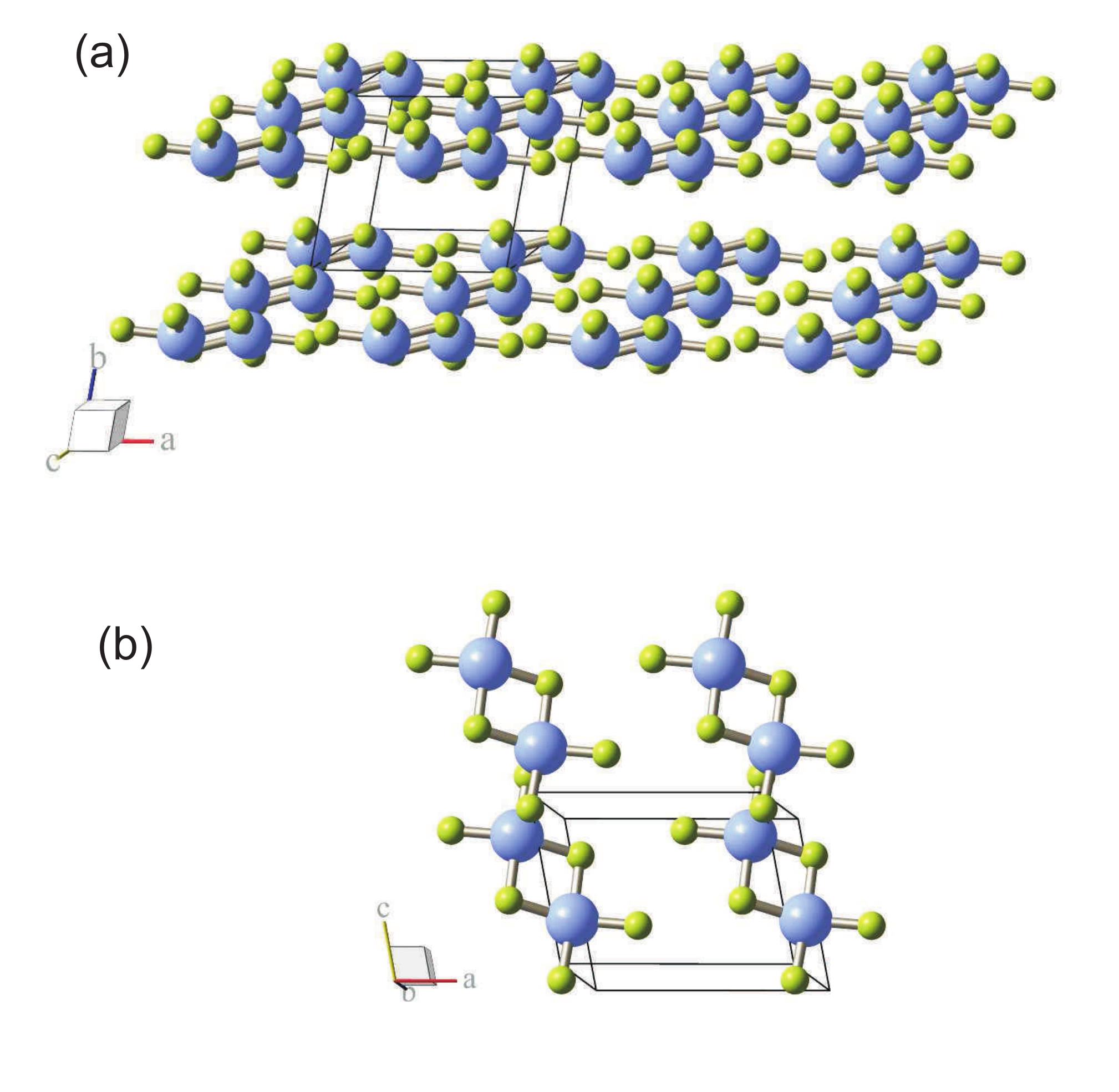}
\caption{(a) Fragment of  \phcc{} crystal structure. Only copper (big balls) and chlorine (small balls) positions are shown, bonds within Cu$_2$Cl$_6$ are drawn for clarity. (b) Fragment of two-dimensional layer within the  $(ac)$ plane. \label{fig:phcc-struct}}
\end{figure}

\begin{figure}
\centering
\includegraphics[width=\figwidth]{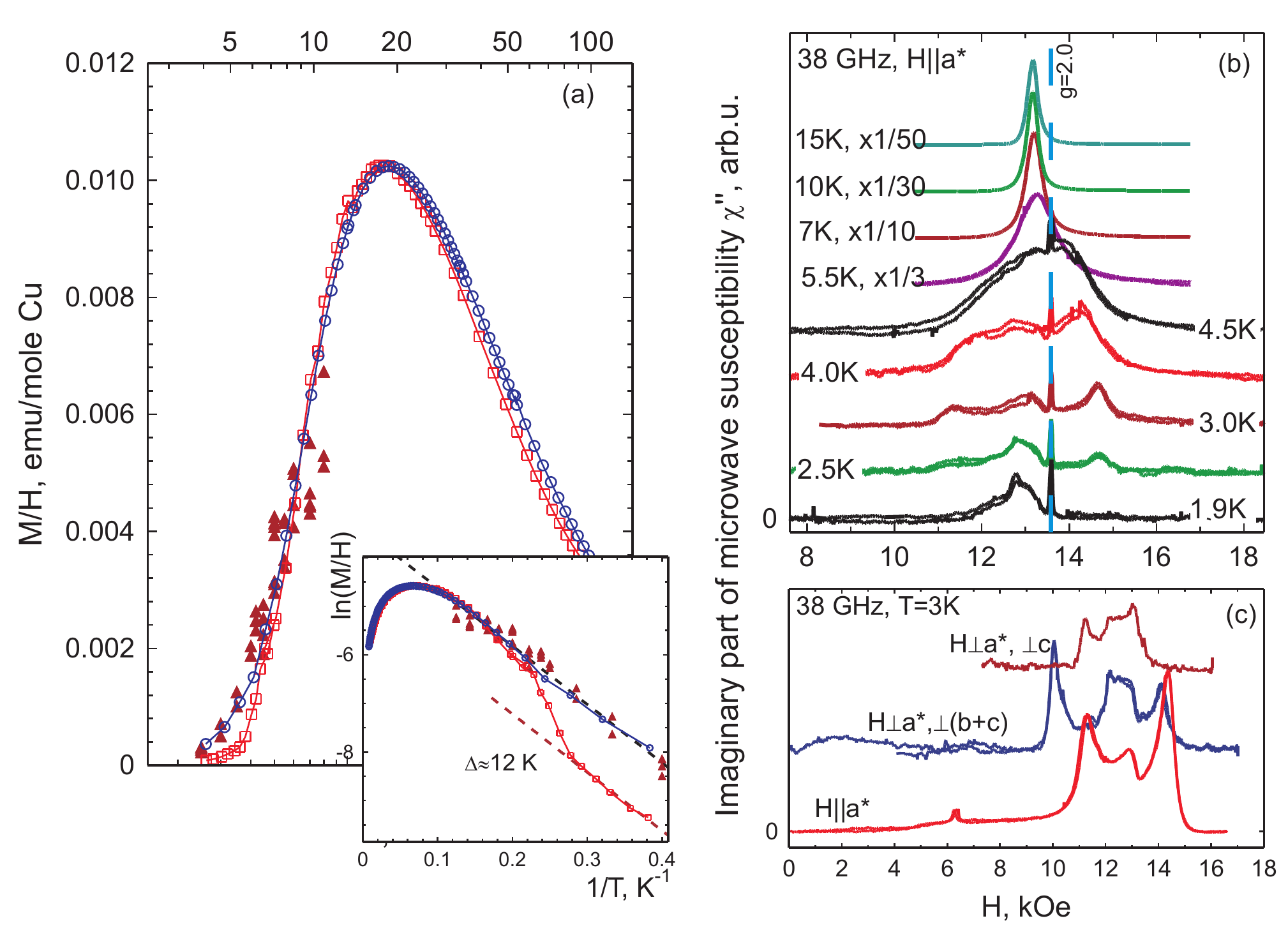}
\caption{(a) Temperature dependence of static susceptibility (circles) and scaled integral intensity of ESR absorption (squares --- X-band data, triangles --- data obtained on microwave frequency of 34~GHz) for  \phcc{}. Inset: dependence of  $\ln(M/H)$ vs. $1/T$, dashed lines corresponds to activation law with activation energy $\Delta=12$~K. (b) Representative ESR absorption curves (recalculated to imaginary part of microwave susceptibility) for \phcc{} at different temperatures $f=38$~GHz, $\vect{H}||a^*$. Data at $T\geq 5.5$~K are scaled for better presentation. Vertical dashed line marks resonance field corresponding to $g=2.0$, narrow absorption line at this resonance field is the DPPH mark.  (c) Comparison of the ESR absorption spectra (imaginary part of microwave susceptibility) at different field orientations. \label{fig:phcc-data1}}
\end{figure}

\begin{figure}
\centering
\includegraphics[width=\figwidth]{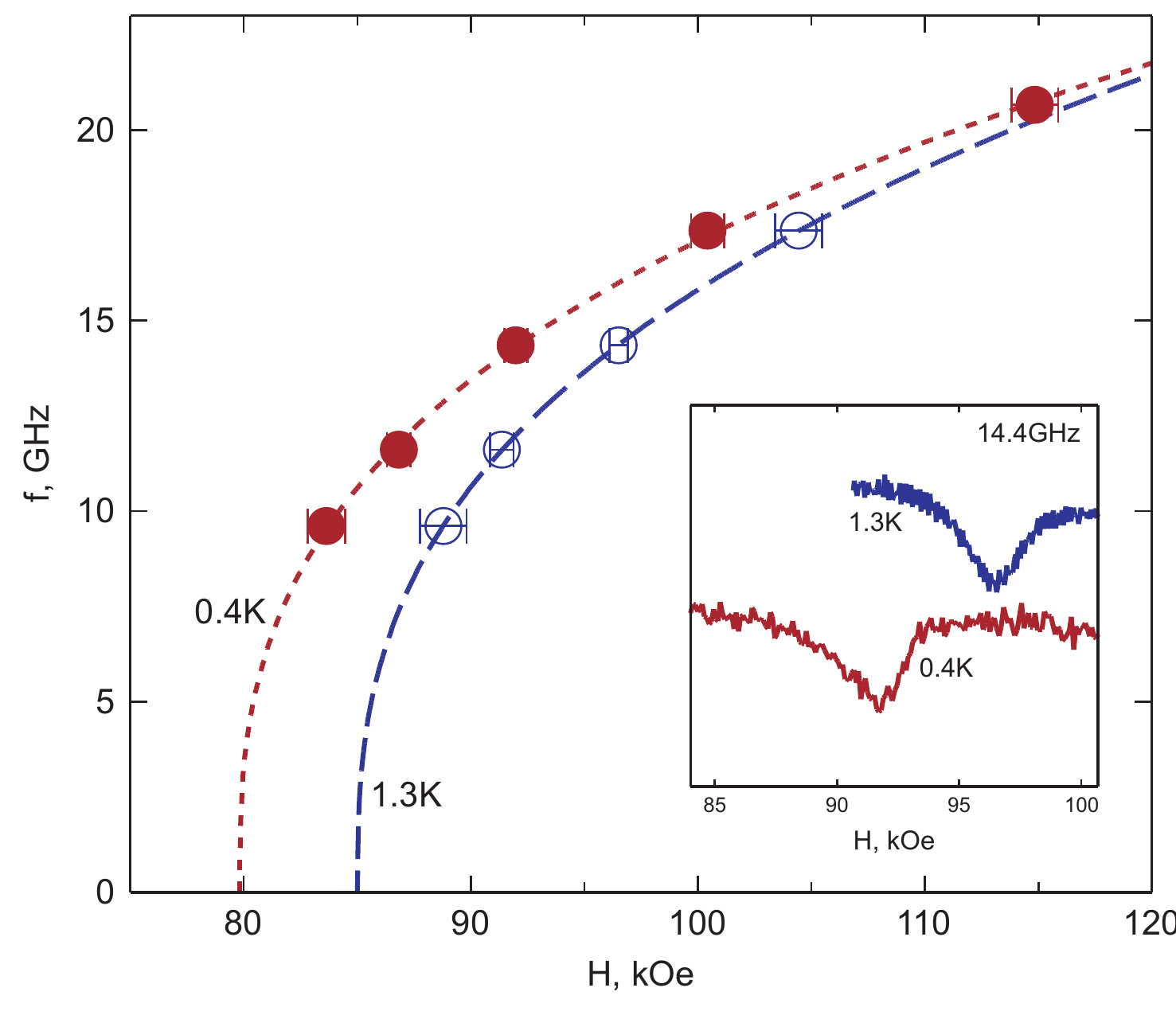}
\caption{Frequency-field diagram for ESR in field induced antiferromagnetically ordered phase of  \phcc{}. Open symbols --- $T=1.3$~K, closed symbols --- $T=0.4$~K. Dashed curves --- fit by  $f\propto\left(H-H_c\right)^\beta$, critical exponents values are given in the text. $\vect{H}||a^*$. Inset: examples of resonance absorption spectra at $f=14.4$~GHz. \label{fig:phcc-afmr}}
\end{figure}

\begin{figure}
\centering
\includegraphics[width=\figwidth]{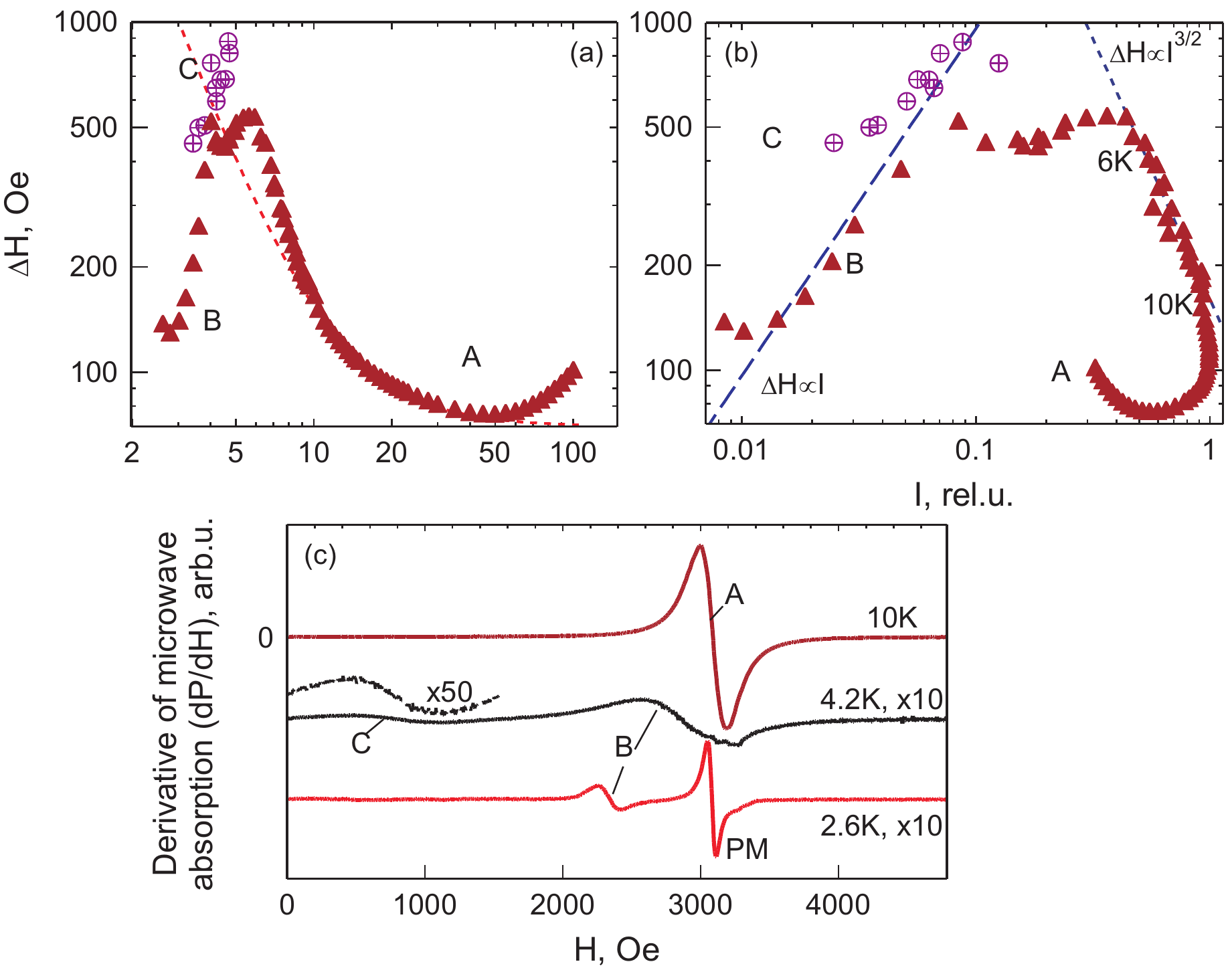}
\caption{(a) Temperature dependence of ESR linewidth in PHCC, letters ``A'', ``B'' and ``C'' mark data for different spectral components. Dashed curve corresponds to phenomenological law $\Delta H=\Delta H_0 (1+(\Theta/T)^2)$ with characteristic temperature $\Theta=11$~K. (b) Linewidth vs. ESR intensity plot for PHCC, letters ``A'', ``B'' and ``C'' mark data for different spectral components. Dashed lines corresponds to the power laws describing fragments of this plot. (c) Examples of ESR absorption field derivative at different temperatures. At 4.2~K and 2.6~K data are scaled for clarity by a factor of  10, low field part of 4.2~K curve is scaled by a factor of 50 to demonstrate low field absorption component more clearly. Letters ``A'', ``B'' and ``C'' mark  different spectral components<, as used on other panels: ``A'' --- ESR absorption with unresolved fine structure, ``B'' --- main component of the resolved fine structure, ``C'' --- low field component. $f=9.4$~GHz, applied field direction is close to $\vect{H}\perp\{ a^*,c\}$. \label{fig:phcc-dh}}
\end{figure}

 Compound \phcc{} (abbreviated as PHCC) is an example quasi-two-dimensional spin-gap magnet. Its magnetic subsystem is built from Cu$_2$Cl$_6$ dimers forming weakly coupled layers within the $(ac)$ planes of triclinic crystal (Fig.\ref{fig:phcc-struct}) \cite{phcc-struct}. Magnetic measurements, thermodynamic properties and inelastic neutron scattering experiments confirm presence of an energy gap $\Delta =1.0$~meV in the excitations spectrum, critical fields are equal to $H_{c1} \simeq 70$~kOe, $H_{c2} \simeq 380$~kOe \cite{broholm-njp2007,broholm-prb64}. Dimers are strongly coupled, measured dispersion of the triplet excitations indicates presence of at least 6 significant exchange bonds. The strongest bond is the intra-dimer one with exchange integral $J_{dim}=1.3$~meV and inter-dimer bonds along the $a$ and $c$ axes with exchange integrals $J_a=0.92$~meV and $J_c=0.3$~meV \cite{broholm-njp2007}.

Above the first critical field $H_{c1}$ PHCC orders antiferromagnetically, as it is directly confirmed by observation of antiferromagnetic Bragg peaks \cite{broholm-njp2007}. Neel temperature at the field of 10~T amounts  2.85~K. Antiferromagnetic ordering was also observed  at zero field under applied hydrostatic pressure $p>p_c \approx 4.3$~kbar \cite{thede-pressurephcc,perren-pressphcc,taohong}.

 High quality single crystals of pure \phcc{} can be grown from the solution \cite{tanya}. As grown samples have volume from several mm$^3$ to $\sim 0.1$~cm$^3$. Samples have natural facets with a well developed face normal to $a^*$ and are elongated along the crystallographic $c$ axis, which facilitate sample mounting within the experimental cell. Residual concentration of paramagnetic defects in nominally pure samples estimated from low temperature magnetic susceptibility curves at $T\ll\Delta$ amounts about 0.05\%.

Thus, PHCC is an example of well characterized quasi-two-dimensional spin-gap magnet, which allows to study dynamics of triplet excitations in a quantum paramagnet state at $H<H_{c1}$, field-induced antiferromagnetic state at $H>H_{c1}$ and the effects of modification of some exchange bonds induced by partial substitution of nonmagnetic ions on magnetic properties of this magnet.

ESR study of the pure PHCC is described in details in Ref. \onlinecite{glazkov-phcc}, we also include in this review new data on magnetic resonance in a field induced antiferromagnetic phase at $H>H_{c1}$. Examples of ESR absorption spectra in pure \phcc{}  are shown on Fig.~\ref{fig:phcc-data1}-b. As temperature decreases below 15~K ESR absorption intensity quickly decrease as the triplet excitations freeze out. ESR linewidth systematically increases on cooling reaching maximum value around 5~K. On further cooling ESR absorption spectrum splits into several components. Unbiased component grows on cooling and corresponds to the ESR absorption by residual defects (magnetic susceptibility at 1.8~K corresponds to  0.04\% of $S=1/2$ paramagnetic centers with respect to the amount of copper ions present). Split components continue to loose intensity on cooling, they correspond to resolved fine structure of $S=1$ ESR in a crystal field. Maximal splitting of the ESR absorption components is about 4~kOe. Observed splitting of the ESR absorption is anisotropic and includes also weak absorption in the field close to half of the resonance field at higher temperatures.

Temperature dependence of ESR integral intensity can be scaled with the temperature dependence of magnetic susceptibility (Fig.\ref{fig:phcc-data1}-a). Low-temperature part of this dependence follows activation law with activation energy 12~K, which is close to the known gap in the excitations spectrum $\Delta=1.0$~meV. Deviation of X-band (9.3~GHz) ESR intensity from this curve followed by approach to similar law on further cooling is explained by the fact that resonance frequency is comparable to the zero-field splitting of triplet sublevels and some of the fine structure components are not observable.

Observed splitting of the ESR absorption spectrum and observation of the weak absorption component at half of the ordinary resonance field are typical for spin $S=1$ in a crystal field \cite{altkoz} and indicate presence of the effective anisotropy, see Eqn.~(\ref{eqn:aniseff}). Low (triclinic) symmetry of PHCC crystal does not put any  constraints on orientation of effective anisotropy axes. To determine anisotropy parameters we have measured angular dependencies of ESR spectra splitting at 3~K, and to determine anisotropy of $g$-factor we have measured angular dependencies of resonance field at 25~K. Joint fit of these dependencies allowed to determine main values of $g$-tensor (which turned to be axial within experiment accuracy) $g_{||}=(2.280\pm0.015)$, $g_{\bot}=(2.048\pm0.007)$ and values of the effective anisotropy constants $D=(-7900\pm280)$~MHz and $E=(-1340\pm190)$~MHz, as well as the orientation of axial  $g$-tensor axis and anisotropy axes with respect to the crystal  \cite{glazkov-phcc}.

We did not observed singlet-triplet ESR transitions in \phcc{}.

At low temperatures we observed ESR absorption signal from the field-induced antiferromagnetic phase of \phcc{}, examples of the resonance absorption curves and frequency-field diagrams are shown in Fig.~\ref{fig:phcc-afmr}. These frequency-field diagrams are analogous to that observed earlier for  TlCuCl$_3$ \cite{tlcucl3-1,tlcucl3-2}: frequency of the low-frequency antiferromagnetic resonance mode for collinear antiferromagnet $f\propto\sqrt{H_A H_E}$ is proportional to the antiferromagnetic order parameter (sublattice magnetization) \cite{gurevich}. Due to low symmetry of PHCC magnetic field orientation $\vect{H}||a^*$  chosen for experiment does not coincide with any of the anisotropy axes which allows coupling of this resonance mode to microwave field. To check relation of the magnetic resonance frequency to the order parameter amplitude \cite{broholm-njp2007} we fitted frequency-field diagrams as

\begin{equation}\label{eqn:phcc-crit}
f\propto \left(H-H_c\right)^\beta,
\end{equation}

\noindent found critical fields values: $(85\pm 1)$~kOe at $T=1.3$~K and $(80\pm 1)$~kOe at $T=0.4$~K agree within error margin with the magnetization data \cite{broholm-njp2007}, and critical exponent $\beta=(0.36\pm0.03)$ coincides with critical exponent value measured from the field dependence of the order parameter found from neutron scattering experiments \cite{broholm-njp2007}.

Measured parameters of the effective anisotropy in low-field paramagnetic phase allow to predict expected order parameter orientation in an ordered phase and to compare this prediction with neutron diffraction results \cite{broholm-njp2007}: it turns out that predicted and measured order parameter orientations differ approximately by 10$^\circ$.

Relation between the anisotropy parameters in low-field and high-field phases can be established within the framework of bosonic (macroscopic) approach \cite{farmar}. Dynamics of spin-gap magnet is described as oscillations of the vector field $\boldeta$, which becomes an order parameter within the ordered phase. Lagrangian of this field can be written as:

\begin{equation}\label{eqn:lagr}
{\cal{L}}=\frac{1}{2} \left(\dot{\boldeta}+\gamma[\boldeta\times
\vect{H}]\right)^2-\frac{1}{2}A\boldeta^2-\beta\boldeta^4+{\cal{L}}_{rel}
\end{equation}
\noindent here $A,\beta>0$ are exchange constants, parameter $A$ determines energy gap, parameter $\beta$ determines order parameter amplitude at $H>H_{c1}=\sqrt{A}/\gamma$. Term ${\cal{L}}_{rel}$ includes anisotropic corrections, we will take into account second order anisotropy and effects related to $g$-factor anisotropy. For second order anisotropy

\begin{equation}\label{eqn:lagr-rel1}
  {\cal{L}}_{rel1}=\frac{1}{2}b_1
  (\boldeta^2-3\eta_{Z_A}^2)+\frac{1}{2}b_2(\eta_{X_A}^2-\eta_{Y_A}^2)
\end{equation}
\noindent here $(X_A,Y_A,Z_A)$ --- anisotropy axes, this form of expression maintains average energy of triplet sublevels constant on changing  $b_{1,2}$. Low symmetry of  PHCC crystal allows numerous symmetry-invariant combinations quadratic on $\boldeta$ components and including magnetic field or time derivative $\dot{\boldeta}$. To describe observed axial anisotropy of $g$-factor we restrict this choice to a combination which ensures linear field dependence of triplet sublevels close to the first critical field in the isotropic limit $b_1=b_2=0$:

\begin{equation}\label{eqn:lagr-rel2}
{\cal{L}}_{rel2}=\xi(\gamma^2 H_{Z_g}^2\boldeta^2-\gamma^2
(\vect{H}\boldeta)H_{Z_g}\eta_{Z_g}+\gamma
H_{Z_g}[\dot{\boldeta}\times\boldeta]_{Z_g})
\end{equation}
 \noindent here $Z_g$ --- is the main axis of the $g$-tensor,  $\xi=(g_{||}/g_{\perp}-1)$.

At zero field eigenfrequencies of the field  $\boldeta$ are:

\begin{equation}\label{eqn:eigenfreqs}
    \omega^2=\left\{\begin{array}{c}
                        A+2b_1 \\
                        A-b_1-b_2 \\
                        A-b_1+b_2 \\
                      \end{array}
                    \right.,
\end{equation}

\noindent which allows easily to establish relation between parameters of perturbative and macroscopic models.

Note that for the simplest axial case with $b_1>0$, $b_2=0$, $\xi=0$ effective anisotropy of the triplet excitations will be of ``easy axis'' type (two-fold degenerate lower sublevels correspond to  $S_z=\pm 1$). In the same time, to search for equilibrium orientation of the order parameter $\boldeta_0$ above $H_{c1}$ one have to minimize potential energy density

\begin{equation}\label{eqn:orderedphase}
    \Pi=-\frac{\gamma^2}{2}\left[\boldeta\times\vect{H}\right]^2+\frac{A}{2}\boldeta^2+\beta\boldeta^4-\frac{b_1}{2}\left(\boldeta^2-3\eta_Z^2\right),
\end{equation}

\noindent which, for the same case of $b_1>0$, corresponds to the ``easy plane'' anisotropy for the order parameter (minimum of anisotropy energy corresponds to $\eta_Z=0$). Thus, a specific \emph{inversion} of the type of anisotropy takes place at the transition to the ordered phase at $H_{c1}$: easy axis of anisotropy for the triplet excitations of paramagnetic low-field phase becomes hard axis of anisotropy for the ordered high-field ($H>H_{c1}$) phase (and, vice versa, for $b_1<0$ hard axis of anisotropy for the paramagnetic phase becomes an easy axis for the order parameter of  the ordered phase). This effect is similar to the result of Ref. \onlinecite{date}: for the Haldane chain built from $S=1$ ions signs of the single-ion anisotropy and effective anisotropy of triplet excitations are opposite.

Temperature dependence of ESR linewidth for \phcc{} is shown at the Fig. \ref{fig:phcc-dh}-a. There is a broad minimum of linewidth at approx. 50~K, increase of the linewidth above this temperature is likely due to activation of spin-phonon relaxation  mechanisms. At lower temperatures ESR linewidth is dominated by spin-spin relaxation processes. Halfwidth at halfheight at 50~K is about 70~Oe. Maximum of the ESR linewidth at temperature about 5~K corresponds to the moment of resolution of fine structure of the triplon ESR absorption spectrum.

Due to existence of numerous relevant exchange bonds in PHCC microscopic problem of determination of  ESR linewidth in high-temperature limit is practically useless. One can note that presence of the inversion center forbids Dzyaloshinskii-Moriya interaction within the dimer and on some of the inter-dimer bonds. Assuming that the main reason of ESR line broadening is the symmetric anisotropic spin-spin interaction we can make a conventional estimate of the linewidth for the exchange-narrowed ESR absorption line:

\begin{equation}\label{eqn:phcc-sae-estimate}
    \Delta H \simeq \frac{G^2}{g\mu_B J},
\end{equation}
\noindent here $G$ is a typical magnitude of symmetric anisotropic coupling and  $J$ is the characteristic exchange integral value. This yields $G \simeq 0.4$~K and $G/J \simeq 0.03$, which is close to the conventional estimate $G/J \sim (\Delta g/g)^2$.

At temperatures from 10~K to 50~K observed temperature dependence of ESR linewidth for PHCC  (Fig.\ref{fig:phcc-dh}-a) is well described by phenomenological quadratic dependence:

\begin{equation}\label{eqn:phcc-HTexp}
\Delta H=\Delta H_0 \left[1+\left(\frac{\Theta}{T}\right)^2\right]
\end{equation}

\noindent with characteristic temperature $\Theta=(11\pm1)$~K. This value of characteristic temperature is close to the typical scale of exchange integrals in \phcc{}, which makes this phenomenological guess a plausible version of the high-temperature expansion for the linewidth. Similar broadening $\Delta H\propto T^{-2}$ was predicted for one-dimensional spin systems at $T\sim J$ in the presence of staggered field \cite{oshiaff,furuya-jpsj}.

On further cooling to approximately  10~K ($T\sim \Delta$) ESR linewidth increases faster on cooling. It reaches maximum at approximately 6~K where absorption spectrum splits and the fine structure develops. At $T<6$~K components of ESR absorption spectrum narrows on cooling as we approach limit of the dilute gas of quasiparticles. To check for possible relation between the spin relaxation at $T<\Delta$ and excitations concentration we plot dependence of the ESR linewidth on intensity of spectral components (Fig.~\ref{fig:phcc-dh}-b). One can see that low-temperature linewidth is proportional to the ESR intensity, i.e. spin relaxation in \phcc{} in this regime is due to two-triplon processes.

At the regime of concentrated gas of triplons (6 to 10~K), when the fine structure of the triplon absorption spectrum is exchange narrowed, we observe power law dependence $\Delta H \propto I^{-3/2}$. We recall here, that the linewidth of exchange-narrowed line is inversely proportional to the effective exchange frequency (i.e., inversely proportional to the frequency of quasiparticles interactions. At high concentration of excitations  sample magnetization (and ESR integral intensity) grows with temperature slower than the  excitations concentration, thus the empiric exponent value above 1 is quite natural. However, the power law and the exact exponent value remains unexplained.

\subsection{Formation of  $S=1$ paramagnetic centers in a spin-gap magnet  \phbc{} with random modulation of exchange bonds}
\begin{figure}
\centering
\includegraphics[width=\figwidth]{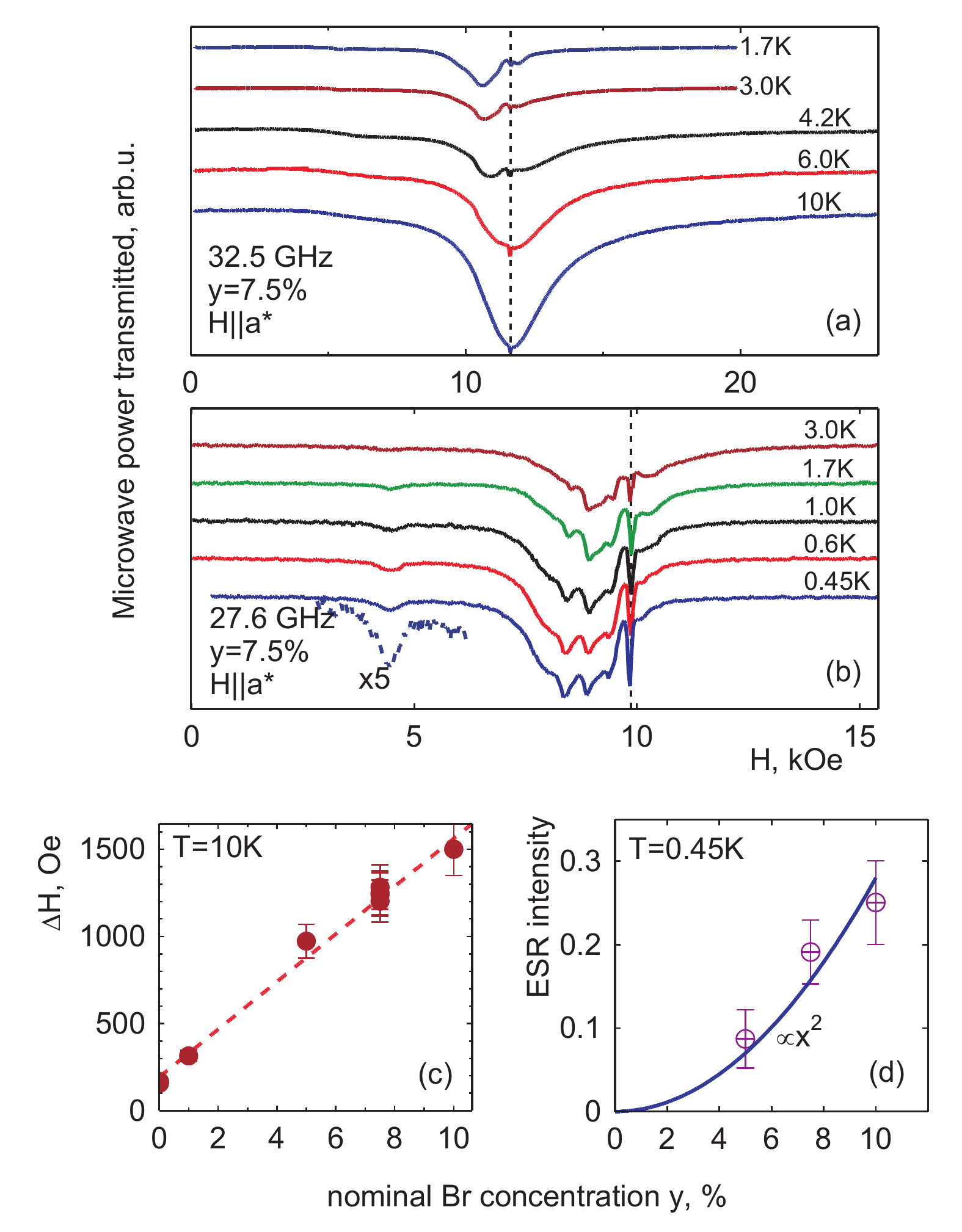}
\caption{(a), (b) Representative examples of ESR absorption spectra in \phbc{} with nominal bromine concentration $y=7.5$\% at different temperatures. Vertical dashed line shows position of DPPH marker ($g=2.00$). Dashed curve at panel (b) --- five-fold Y-magnified fragment of 0.45~K absorption spectrum. (c) Dependence of ESR halfwidth at halfheight on nominal bromine concentration at $T=10$~K. (d) Dependence of the scaled ESR absorption intensity on nominal bromine concentration at $T=0.45$~K. All data are taken at $\vect{H}||a^*$. \label{fig:phcc-doped}}
\end{figure}

It is possible to grow isostructural compound \phbc{} with up to 12\% of chlorine ions substituted by bromine (here $y$ is the nominal concentration of bromine in a growth solution). Inelastic neutron scattering, magnetization and specific heat measurements prove that this substitution does not violate singlet ground state and the gap in the excitations spectrum even slightly increases on chlorine/bromine dilution \cite{dan1,dan2,dan3}. Structural analysis \cite{dan1} demonstrated that the real bromine concentration is slightly less than the nominal concentration in the growth solution (real bromine concentration averaged over three inequivalent position is  $x=0.63 y$) and is slightly different for different chlorine positions  \cite{dan2}. Nonmagnetic chlorine ions in \phcc{} structure are responsible for the formation of superexchange bonds between the magnetic copper ions (Fig.\ref{fig:phcc-struct}). Thus, substitution of chlorine ions by bromine ions results in the appearance of the modified exchange bonds randomly distributed in the crystal.  ESR on bromine diluted \phcc{} was studied in details in Refs. \onlinecite{glazkov-phcc-doped1,glazkov-phcc-doped2,glazkov-phcc-doped3}.

Examples of ESR absorption in bromine-diluted PHCC are shown at Fig. \ref{fig:phcc-doped}. On cooling below approx. 10~K intensity of ESR absorption freeze out indicating presence of an energy gap. Compared to pure PHCC, bromine-diluted \phbc{} demonstrate higher linewidth, within experiment accuracy linewidth increases linearly on bromine content. This means that each bromine ion creates additional relaxation center for triplet excitations. This observation confirms broadening of triplet level with bromine dilution reported in inelastic neutron scattering experiment \cite{dan2}.

However, on cooling below 1~K ESR absorption in  \phbc{} demonstrates characteristic featured of $S=1$ paramagnetic centers. The most clear fingerprint is the ``two-quantum'' transition observed in the magnetic field close to the half of the field of  ordinary magnetic resonance absorption. The main resonance absorption signal at low temperatures is shifted from its position at $T\simeq 10$~K indicating presence of zero-field splitting of energy sublevels. These absorption signals gain intensity on cooling and remain visible down to the lowest temperature of our experiment 0.45~K, within the accuracy of our experiment we can conclude that these ESR absorption signals are not thermally activated. Thus, partial substitution of chlorine ions mediating the Cu-Cu superexchange pathes in PHCC by bromine ions results in the formation of paramagnetic centers with spin  $S=1$. Effective crystal field acting on these centers is different from the effective crystal field acting on triplet excitations of pure compound: magnitude of fine structure splitting in \phbc{} is slightly less and signs of the effective anisotropy constant measured at $\vect{H}||a^{*}$ in pure and bromine diluted compounds are opposite. Similar effect of formation of  $S=1$ paramagnetic centers in a  $S=1/2$ molecular magnet of ``spin ladder'' type [Ph(NH$_3$)]([18]crown-6)[Ni(dmit)$_2$] was reported in Ref.~\onlinecite{fuji-ladder-s1}.

Static magnetization measurements demonstrated that for nominal bromine concentration $y=10$\% magnetization at  $T=1.8$~K equals to that of approx. 1.5\% of paramagnetic centers (as compared to the amount of copper ions) with spin $S=1/2$ \cite{glazkov-phcc-doped3}. ESR spectroscopy proves that magnetic centers responsible for low temperature magnetic properties have spin $S=1$. Comparison of the ESR intensity temperature dependence measured down to 0.45~K with static magnetization data above 2~K (measured with SQUID-magnetometer on the samples from the same batches) allowed to scale ESR intensity measured on different samples with different bromine content  to the same absolute units. Concentration dependence of the scaled ESR intensity at 0.45~K is shown at Fig. \ref{fig:phcc-doped}. This intensity, which is directly proportional to the concentration of $S=1$ centers, grows quadratically on bromine concentration.

Concentration of $S=1$ centers can be calculated from the scaled ESR intensity data using Curie law. It turned out that for nominal bromine concentration $y=10$\% thus determined actual concentration of $S=1$ centers amount to $(0.4\pm0.1)$\% per molecule of \phcc{} (or $(0.20\pm0.05)$\% per copper ion). We suggest that $S=1$ center appears on simultaneous substitution of two intra-dimer (position Cl$_1$ in the notations of Ref.~\onlinecite{dan2}) chlorine ions by bromine ions. Real bromine occupation of these intra-dimer positions at  $y=10$\% equals 7.5\% \cite{dan2}, thus probability of such a double substitution within the dimer is $(0.075)^2=0.56$\% --- which is really close to the measured value. This model is also supported by the details of PHCC structure: intra-dimer Cu-Cl-Cu bond angle is equal to 95.8$^\circ$ (i.e., it is close to 90$^\circ$), hence even small changes of this angle by the chemical pressure on halogen substitution could strongly affect corresponding exchange coupling constant. Study of the excitations spectra in PHCC under hydrostatic pressure \cite{perren-pressphcc} also demonstrated that the same intra-dimer bond is the most affected by the applied hydrostatic pressure.

Effect of bromine substitution on the properties of PHCC can be qualitatively described as follows. At low bromine content one can assume that modifications of exchange bonds created by bromine substitution are equivalent to some scattering potential for triplons with scattering centers randomly distributed through the PHCC matrix. The inelastic (accompanied by spin relaxation) scattering by this potential causes linear increase of ESR linewidth with increasing bromine content. Possible reason for such effective inelastic scattering is that substitution of single chlorine ion on any exchange bond violates inversion symmetry and hence allows for Dzyaloshinskii-Moriya interaction on this bond, which affects the spin relaxation. In the same time, being averaged, this random potential results in the increase of the energy of triplons able to travel through the crystal, i.e., it leads to the increase of the gap, as observed in \phbc{} \cite{dan2}. Localization of triplons in this random potential requires separate study: since network of exchange bonds in PHCC is 3D it is possible that potential wells created by bromine substitution are too shallow to produce  a bound state. However, our ESR experiment indicate that double substitution of halogens on the same bond does create a deep potential well with bound triplon state, which becomes the $S=1$ center. In the extreme limit this localized triplon could be imagined microscopically as a formation of strong  ferromagnetic bond in a doubly substituted Cu$_2$Cl$_6$ dimer.

\subsection{Quasi-one-dimensional ``spin-tube'' magnet \sul{}: singlet-triplet transitions and helical ordering}
\begin{figure}
\centering
\includegraphics[width=\figwidth]{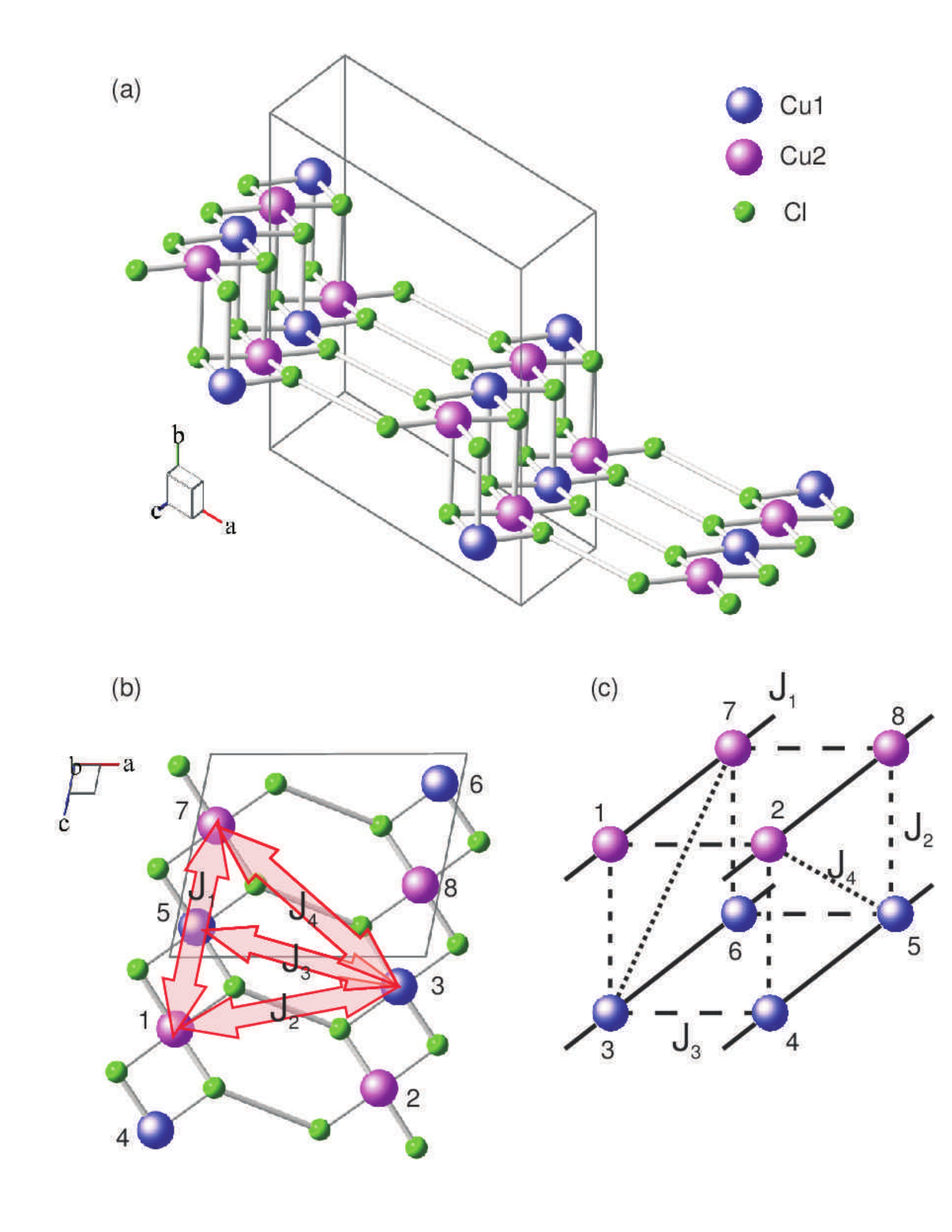}
\caption{Fragment of \sul{} crystal structure, only positions of copper and chlorine ions are shown for clarity. Nonequivalent copper positions are marked by different shades of color. (a) General view of crystal structure. (b) 1D fragment of crystal structure with the main exchange bonds  $J_{1,2,3,4}$ (c) Equivalent representation of ``spin tube'', notation of copper ions and exchange bonds is the same as in panel (b). \label{fig:sul-struct}}
\end{figure}
\begin{figure}
\centering
\includegraphics[width=\figwidth]{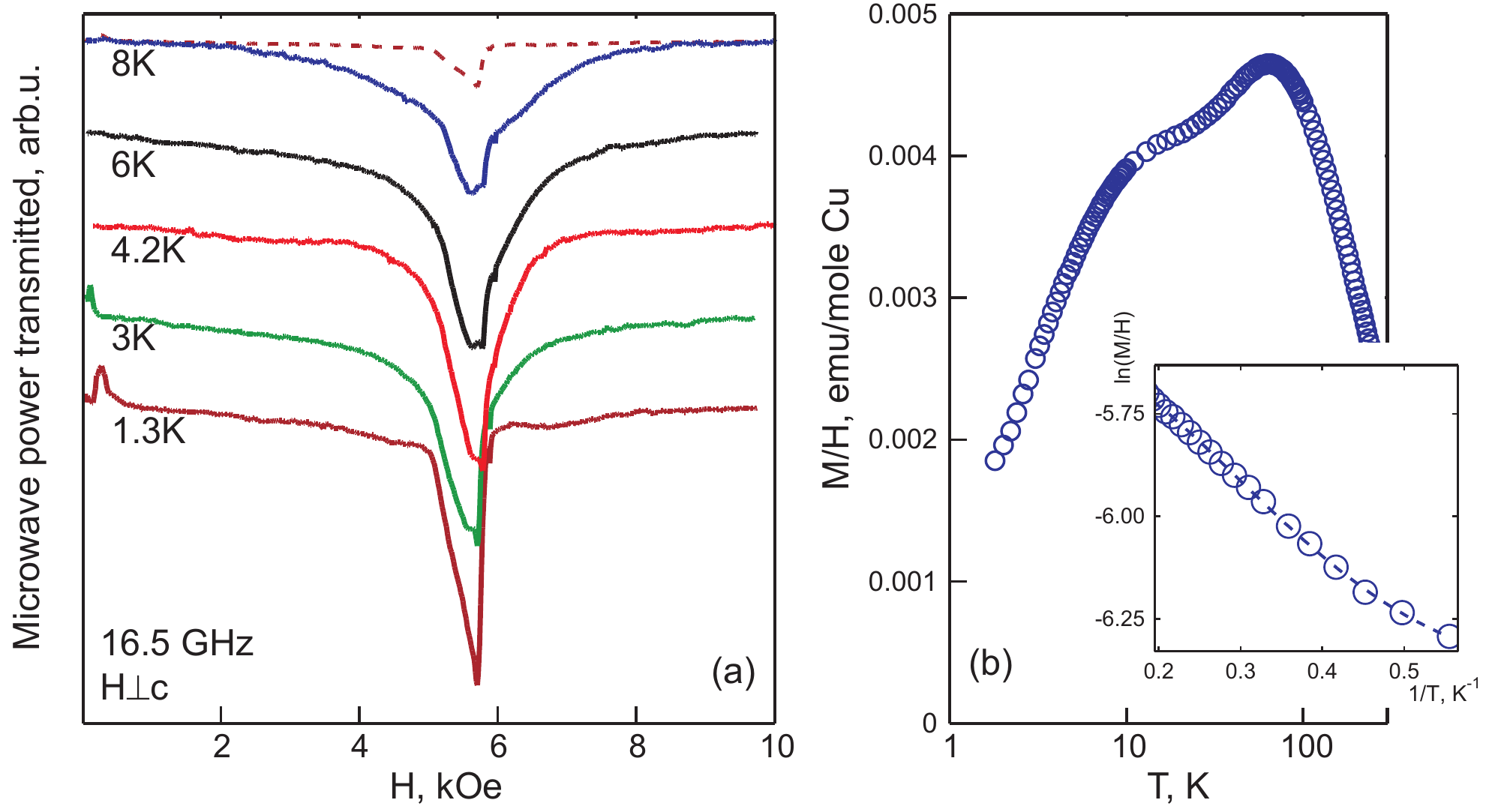}
\caption{(a) Examples of ESR absorption spectra in \sul{} at temperatures from  1.3~K to 8~K. $f=16.5$~GHz, $\vect{H}$ is orthogonal to the sample plane. Dashed curve shows for comparison ESR absorption at 1.3~K scaled by a factor $(1.3/8)$ according to Curie law. (b) Temperature dependence of magnetic susceptibility for \sul{}. Inset --- $ln(M/H)$ vs. $1/T$ plot, dashed curve is a fit by a sum of 1D spin-gap magnet contribution and Curie contribution from the contaminated sample surface $M/H=a/T+(b/\sqrt{T}) e^{-\Delta/T}$ with activation energy $\Delta=4.9$~K. \label{fig:sul-spectra}}
\end{figure}
\begin{figure}
\centering
\includegraphics[width=\figwidth]{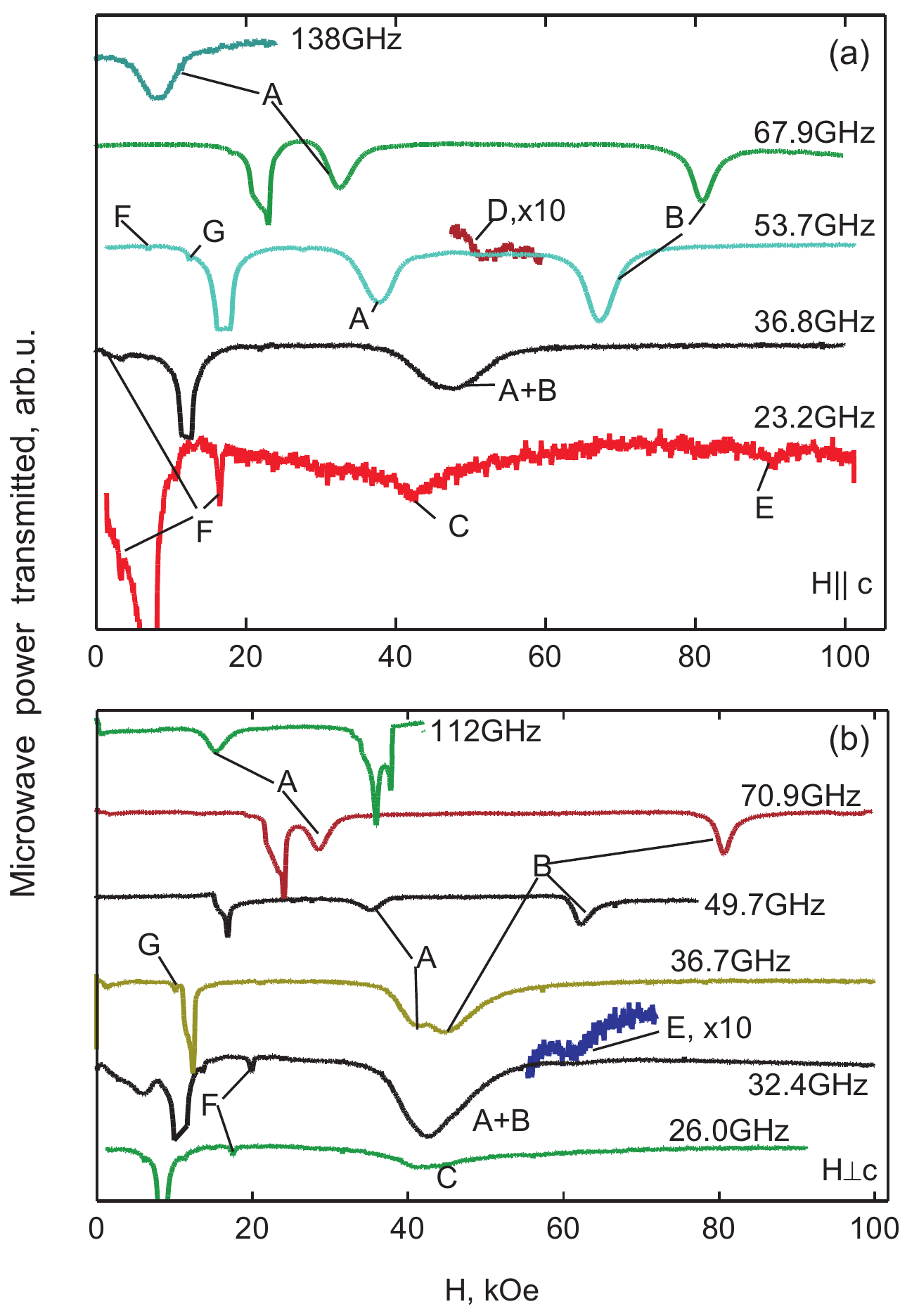}
\caption{Examples of ESR absorption spectra in \sul{} at 0.45~K at various microwave frequencies. (a) $\vect{H}||c$ (along the chains), (b) $\vect{H}$ orthogonal to the sample plane \label{fig:sul-spectra(f)}}
\end{figure}
\begin{figure}
\centering
\includegraphics[width=\figwidth]{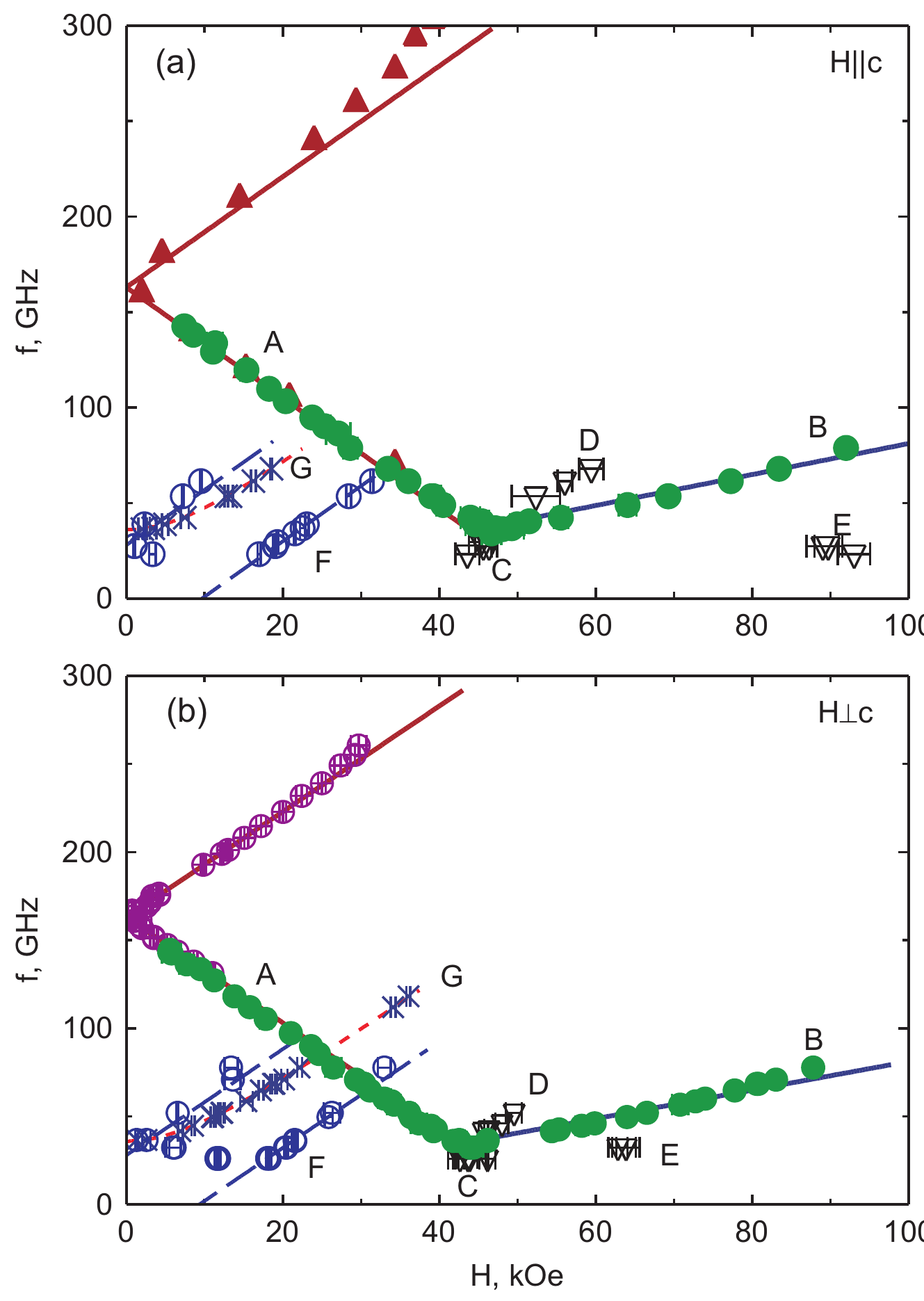}
\caption{Frequency-field diagrams for \sul{} ((a) $\vect{H}||c$ (along the chains), (b) $\vect{H}$ orthogonal to the sample plane). Notations of ESR modes is the same as in  Fig.\ref{fig:sul-spectra(f)}. Data at $f<140$~GHz are taken at $T=0.45$~K, data at $f>140$~GHz in panel (b) are taken at $T=1.3$~K, high-frequency data in panel (a) are from M.Fujisawa paper  \cite{fuji-esr}. Solid lines are model $f(H)$ dependencies for the main ESR modes as described in the text, dashed curves --- $f(H)$ dependencies for the weaker modes ``F'' and ``G'' as described in the text. \label{fig:sul-f(h)}}
\end{figure}

Metal-organic compound \sul{} (also named  sul-Cu${2}$Cl$_4$ in the literature) us an unique example of exchange bond geometry of the ``spin-tube'' type. Structure of this compound  \cite{fuji-sruct-jpsj} is shown in Fig.~\ref{fig:sul-struct}, it includes two inequivalent copper ion positions which form structural elements looking like a ladder. However,  Cu-Cl-Cu bond angles between nearest copper ions are close to  90$^\circ$ which significantly weakens coupling between nearest ions. Superexchange bonds via two intermediate halogens become more important (Fig.~\ref{fig:sul-struct}-b) and the resulting geometry of the exchange bonds can be envisioned as a ``spin tube'': four parallel spin chains are coupled by inter-chain interactions (Fig.~\ref{fig:sul-struct}-c) \cite{zhelud-sul2,zhelud-sul3}. Low symmetry of \sul{} crystal ($P\overline{1}$) does not put strong constraints on the possible anisotropic interactions. One can note that bond $J_1$ along the ``spin tube'' is created by translation, hence Dzyaloshinskii-Moriya interaction uniform along the legs of ``spin tube'' is allowed. Inversion center in the middle of transverse bond $J_3$ forbids Dzyaloshinskii-Moriya interaction there. The same inversion center requires that directions of Dzyaloshinskii-Moriya vectors are exactly opposite on the legs of the ``spin tube'' including ions (see Fig.~\ref{fig:sul-struct})  $(1;7)$ and $(2;8)$, and on the legs including ions  $(3;6)$ and $(4;5)$. Since inequivalence of copper ions  and their halogen cages is very minute, one can reasonably assume that Dzyaloshinskii-Moriya vectors are almost parallel on the legs including ions $(1;7)$ and $(4;5)$ (the same for the legs including ions $(2;8)$ and $(3;6)$). Dzyaloshinskii-Moriya coupling was discussed as a possible reason of the allowed  singlet-triplet transitions in \sul{} \cite{fuji-esr}. In the same time, Ref.~\onlinecite{furuya-jpsj} suggested that spin relaxation in this compound is determined by symmetric anisotropic interactions.

Temperature dependence of magnetization (Fig.~\ref{fig:sul-spectra}-b) demonstrate a broad maximum at temperatures around 65~K, which is a characteristic feature of low-dimensional magnets and indicates presence of exchange couplings with typical scale of exchange integral $J\simeq 100$~K \cite{fuji-sruct-jpsj}. At temperatures below 10~K magnetic susceptibility quickly freeze out on cooling, pointing to the gap in the excitations spectrum. The gap value according to inelastic neutron scattering data  \cite{zhelud-sul2} equals $\Delta=0.5$~meV. Temperature dependence of susceptibility has a very narrow temperature range for comparison with thermal activation law since low temperature magnetization is strongly affected by contribution of paramagnetic defects. However, from 2~K to 5~K temperature dependence of magnetization can be described as a sum of Curie contribution and activation law for one-dimensional spin-gap magnet (Eqn.~(\ref{eqn:chi(T)})): $M/H=a/T+(b/\sqrt{T}) e^{-\Delta/T}$ with activation energy $\Delta=4.9$~K. Energy gap can be closed by applied magnetic field of approx. 4~T, at the field of 55~T magnetization reaches about  30\% of saturation value \cite{fuji-sruct-jpsj}.

Transition to the field-induced antiferromagnetic state was observed in  \sul{} both as lambda-anomalies on specific heat curves  \cite{fuji-progrtheory,fuji-disser} and as rising magnetic Bragg peaks above $H_{c1}$ \cite{zhelud-sul1}. Ordering temperature in the field of 10~T is about 1.5~K \cite{fuji-progrtheory}. Accurate analysis revealed that Bragg peaks are located at the incommensurate wavevector $\vect{k}=(-0.22,0, 0.48)$ \cite{zhelud-sul1}, excitations spectrum of the ordered phase includes gapless Goldstone mode typical for helical magnets  corresponding to the arbitrary translation along the helix \cite{zhelud-sul2}. Above the critical field \sul{} demonstrates multiferroic properties \cite{sul-multiferr}, which is also related to helical magnetic structure. Additionally, high-resolution measurements of excitations spectrum in low field paramagnetic phase revealed that minimum of excitations spectrum is shifted from typical for 1D spin-gap magnet antiferromagnetic wavevector to incommensurate position: $q_{min} c=\pi-0.044$ \cite{zhelud-sul3}.

Magnetic resonance in \sul{} was studied by M.Fujisawa  \cite{fuji-esr,fuji-disser}. These studies revealed presence of singlet-triplet transitions and determined ESR linewidth at high frequencies (above 100~GHz) from 2~K to 300~K. Temperature dependence of linewidth was analyzed in Ref.~\onlinecite{furuya-jpsj} within framework of spin-ladder model. Here we will describe new results including observation of resonance modes in the field induced antiferromagnetic phase at temperatures down to 0.45~K.

Samples of  \sul{} were grown as described in Ref.~\onlinecite{tanya}, we used samples with the mass up to 100~mg in our experiments. Samples were unstable under ambient conditions, to reduce samples deterioration we sealed mounted samples with paraffine. However, samples surface was always contaminated by paramagnetic defects. This does not allowed to study in details fine structure of triplet excitations of ESR spectra and spin relaxation in this compound. As grown samples had a developed plane with long edge along the ``spin tubes'' (crystallographic $c$ axis) direction. We have carried out our measurements in two orientations convenient for sample mounting: for the field applied along the ``spin tubes'' (along the $c$-axis) and for the field applied orthogonal to the developed plan, these directions are named  $H||c$ and $H\perp c$ in the Figures for short.

Examples of ESR absorption spectra and temperature dependence of magnetization are shown in Fig.~\ref{fig:sul-spectra}. We observe broad ESR absorption line with halfwidth about 1~kOe at  $T=8$~K, which is close to the values reported in high-frequency ESR study \cite{fuji-esr}, and irregularly shaped paramagnetic absorption line from the parasitic phase on the deteriorated sample surface. As temperature decreases main (broad) absorption signal freeze out, while irregular line increases its intensity approximately following the Curie law. This allows to ascribe observed broad absorption line to triplet excitations, but its big linewidth and closely located parasitic absorption signal make it inconvenient for the detailed study.  However there are other resonance modes well separated from the parasitic absorption signal (Fig.~\ref{fig:sul-spectra(f)}). First, on cooling down to 1.7~K intense resonance mode marked as ``A'' appears: its frequency at zero field is about 150~GHz, at lower frequencies resonance field of this mode increases as frequency decreases while at higher frequencies resonance field increases as frequency increases. On cooling to the base temperature of 0.45~K new mode ``B'' arises above the critical field $H_{c1}$, which is another example of the antiferromagnetic resonance in the field-induced ordered phase. Modes ``A'' and ``B'' merges at the frequency of approx. 30~GHz, at lower frequencies single absorption signal named as ``C'' is observed. Intensity of the ``C'' mode seems to be weaker as compared to the intensity of modes ``A'' and ``B'' at slightly higher frequencies, however this difference is hard to quantify for multi-mode ESR spectrometer, especially for non-paramagnetic resonance modes with probably unusual excitation conditions. Besides of these main modes we observed some weak modes both above the critical field (``D'' and ``E'') and around the paramagnetic absorption signal (``F'' and ``G'').

Frequency-field diagrams for all modes for two field orientations are shown in Fig.~ \ref{fig:sul-f(h)}.

One can see that mode ``A'' unambiguously corresponds to the singlet-triplet transition, we observe transitions both on the falling and rising triplet sublevels. Slope of this mode corresponds to $g$-factor values equal to $(2.07\pm0.05)$ for $H||c$ and $(2.14\pm0.02)$ for $H\perp c$. Zero field resonance frequency $f_0=(163\pm3)$~GHz, within experiment precision we do not observe any difference in the value of $f_0$ for $H||c$ and $H\perp c$ (such difference is expected if sufficiently strong zero-field splitting of the triplet sublevels is present). Value of the energy gap in \sul{} known from inelastic neutron scattering experiments is  $\Delta=0.5$~meV, which corresponds to 121~GHz in frequency units. This value is remarkably smaller than the found $f_0$ value. This means, that ESR active singlet-triplet transition mixes states away from excitations spectrum minimum, which is in agreement with incommensurate location of the bottom of excitations spectrum in  \sul{} \cite{zhelud-sul3}. Quite probably there are some staggered interactions in \sul{} allowing for ESR-active singlet-triplet transitions at antiferromagnetic wavevector.

Change of the slope on transition from mode ``A'' to mode ``B'' corresponds to the transition over the critical field $H_{c1}$ equal to $(47\pm 2)$~kOe for $\vect{H}||c$ and  $(44\pm2)$~kOe for the field applied orthogonal to the sample plane. Minimal frequency of magnetic resonance at $H_{c1}$ is  $(30\pm 2)$~GHz. Mode ``B'' corresponds to one of the antiferromagnetic resonance eigenfrequencies of the ordered phase.

In the case of noncollinear helical magnetic ordering one expects three modes of magnetic resonance \cite{andmar,svistfar}: one mode approaches Larmor frequency, another mode has zero frequency (it corresponds to the Goldstone mode) and the last mode under condition $\chi_\perp>\chi_{||}$ (here $\chi_\perp$ and $\chi_{||}$  are magnetic susceptibilities of the helical structure for the field applied orthogonal to the plane of the helix and parallel to the plane of the helix, correspondingly) has linear asymptote  $\omega=\gamma H \sqrt{\frac{\chi_{\perp}}{\chi_{||}}-1}$ \cite{svistfar}. Our data (Fig.~\ref{fig:sul-f(h)}) at $H>H_{c1}$, indeed, follows this dependence, slope of the mode ``B'' is practically the same for both field directions studied and corresponds to $\sqrt{\frac{\chi_{\perp}}{\chi_{||}}-1}=(0.29\pm0.02)$ ($\frac{\chi_{\perp}}{\chi_{||}}\approx 1.08$).

Origin of the weak high-field modes ``C'',, ``D'' and ``E'' remains unclear. Mode ``F'' observed at low fields has very narrow absorption signal which does not look like signal from randomly oriented surface defects signal (one expect powder averaged signal from the deteriorated sample surface). Frequency-field dependence for the mode ``F'' is similar to that for $S=1$ in a crystal field  \cite{altkoz}. Positions of these resonances (see Fig.~\ref{fig:sul-f(h)}) is mostly described by linear law $f=g\mu_B B/(2\pi\hbar)\pm\Delta_F$ with $\Delta_F=(28\pm2)$~GHz. There is an intriguing coincidence, probably indicating that this mode is related to some of the eigenmodes of \sul{}, that zero-field frequency $\Delta_F$ is very close to the minimal frequency of modes ``A'' and ``B'' at $H_{c1}$. Mode ``G'' with nonlinear frequency-field dependence with the gap of approx. 36~GHz and temperature dependence of resonance field typical for convenient antiferromagnet was identified as antiferromagnetic resonance signal from  CuCl$_2 \cdot$2H$_2$O, probably formed on deteriorated sample surface under ambient conditions.

\subsection{Quasi-one-dimensional ``spin-ladder'' magnet DIMPY (\dimpy): regimes of spin relaxation at various temperatures}

\begin{figure}
\centering
\includegraphics[width=\figwidth]{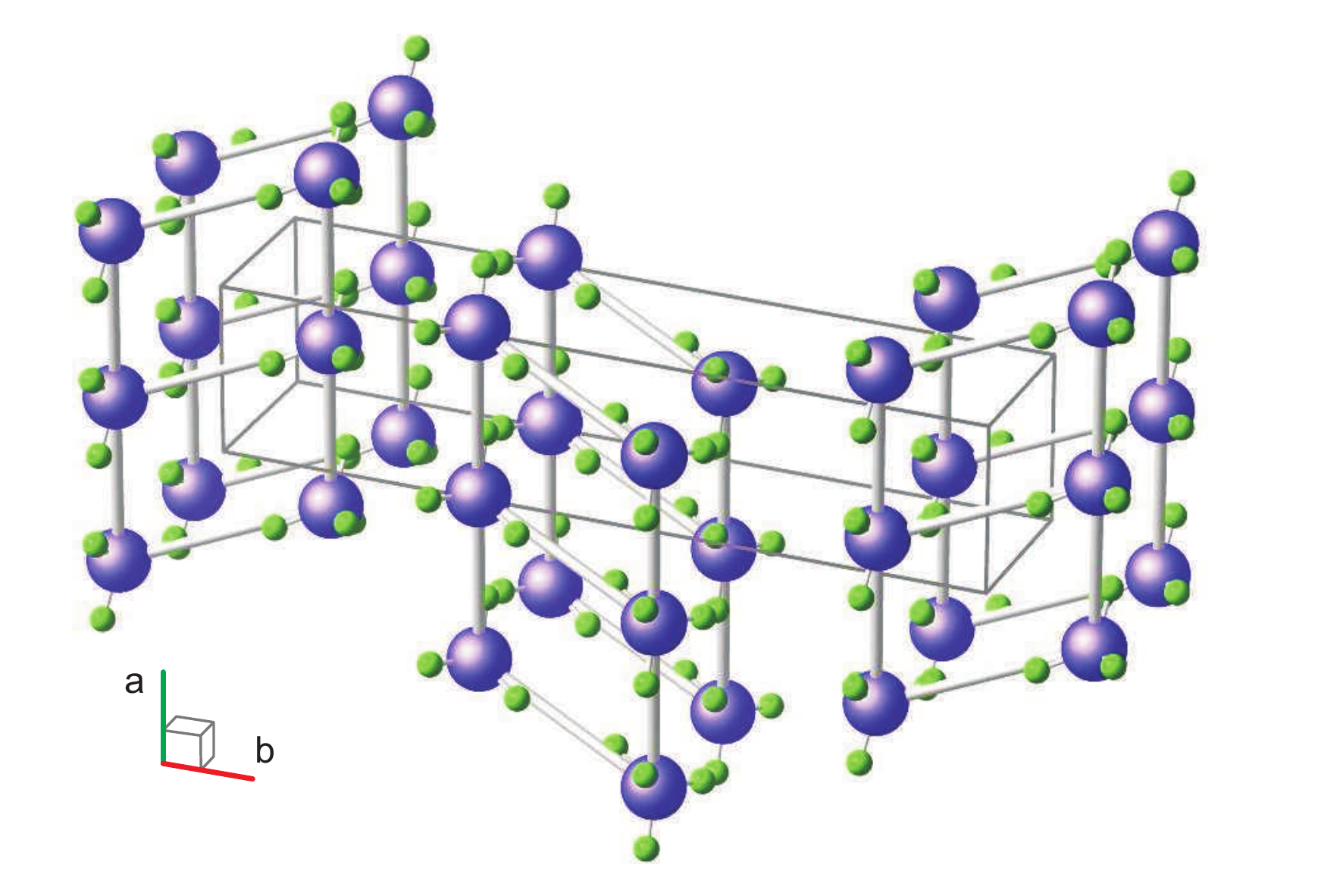}
\caption{Fragment of the crystal structure of quasi-one-dimensional spin-ladder magnet  \dimpy{}. Only positions of copper  (big blue balls) and bromine (small green balls) ions, as well as the exchange bonds forming ladder structures, are shown for clarity. \label{fig:dimpy-struct}}
\end{figure}

\begin{figure}
\centering
\includegraphics[width=\figwidth]{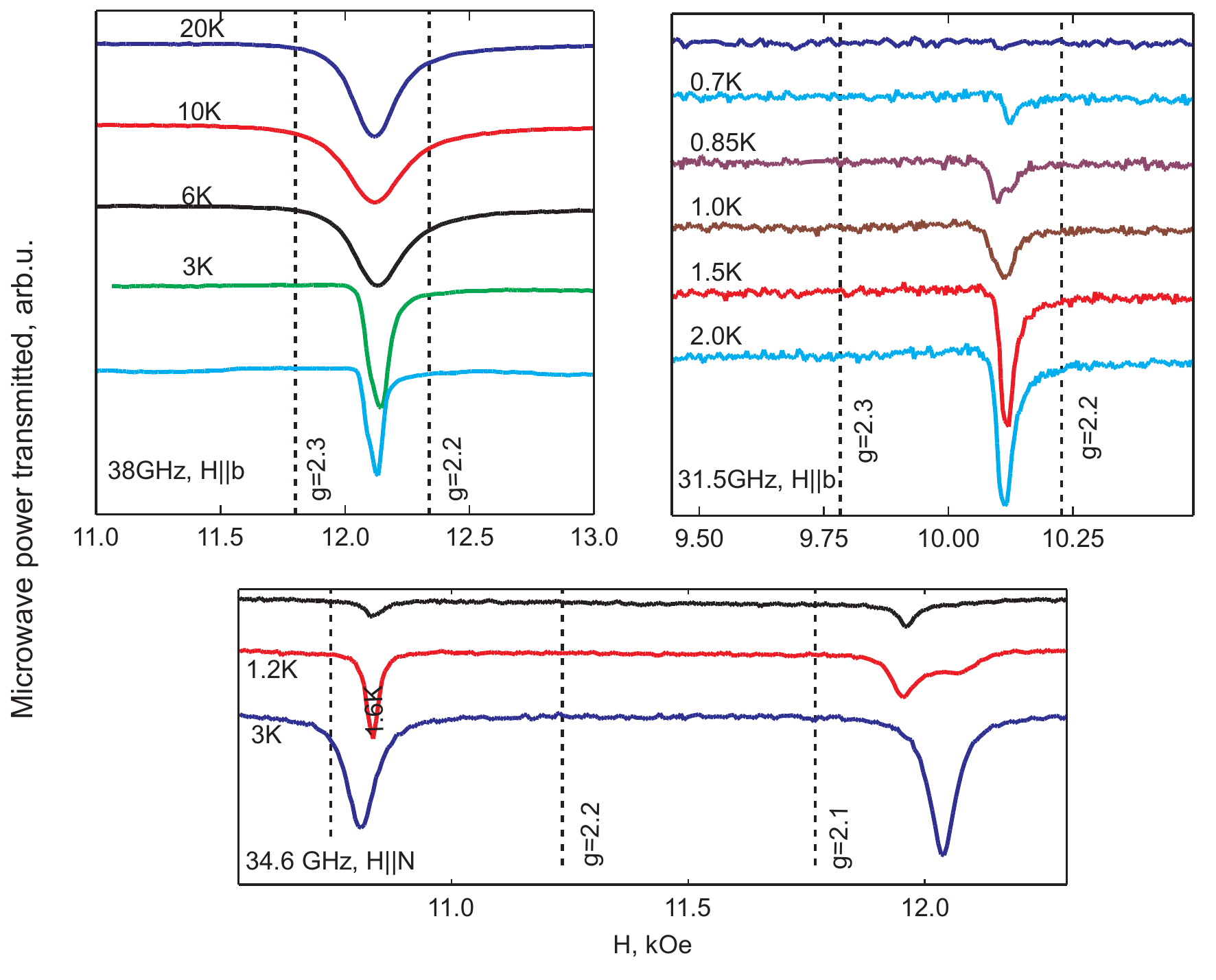}
\caption{Examples of ESR absorption spectra for  \dimpy{} at various temperatures and applied field orientations. Vertical dashed lines mark resonance field positions corresponding to the given $g$-values. \label{fig:dimpy-spectr}}
\end{figure}
\begin{figure}
\centering
\includegraphics[width=\figwidth]{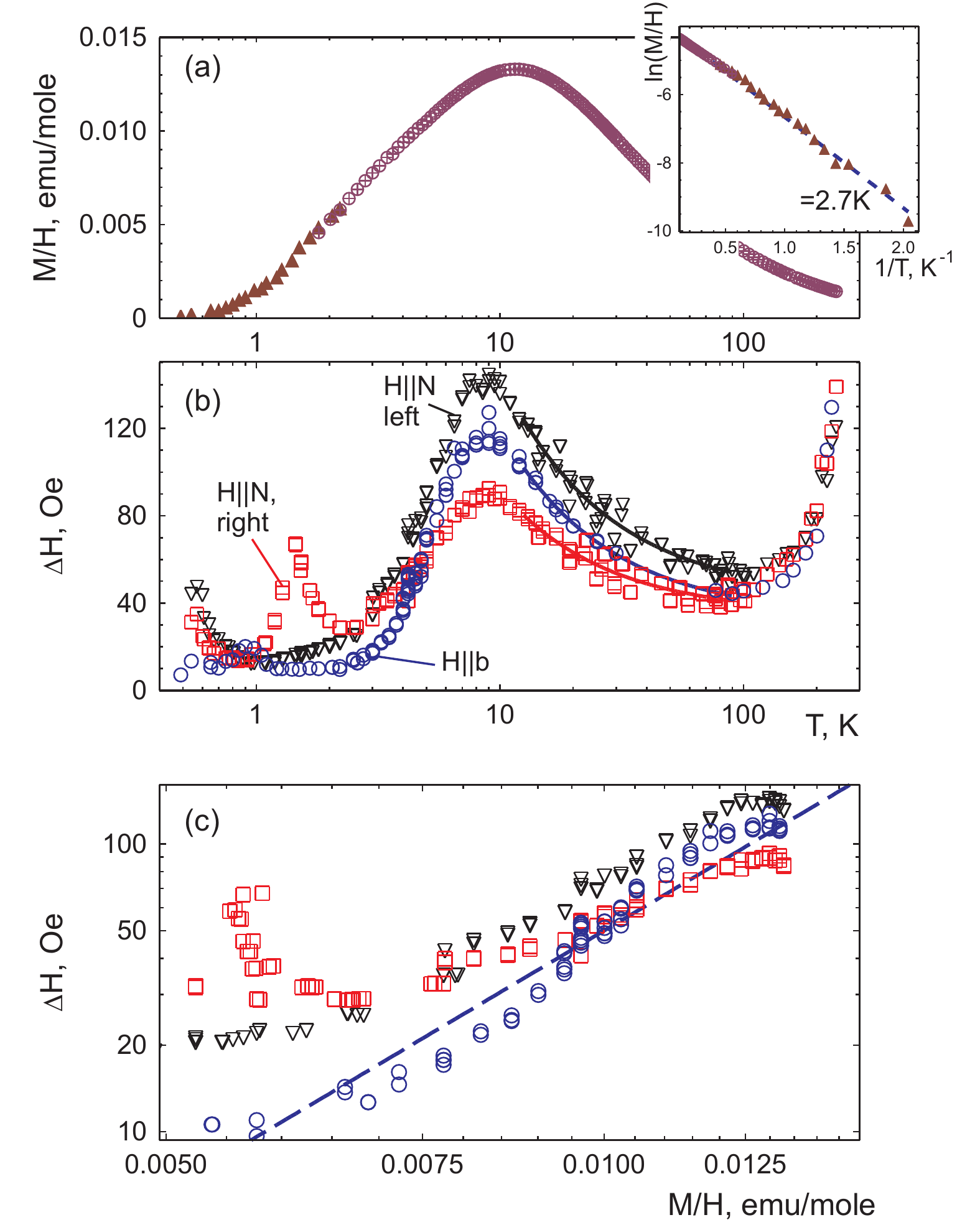}
\caption{(a) Temperature dependence of the magnetic susceptibility of \dimpy{}. Data above 2~K are measured with Quantum Design MPMS SQUID magnetometer, data below 2~K show scaled integral intensity of ESR absorption. Inset:  $ln(M/H)$ vs. $1/T$ plot, dashed line corresponds to activation energy $\Delta=2.7$~K. (b) Temperature dependence of ESR linewidth (halfwidth at halfheight) at different sample orientations, for $\vect{H}||N$ data for two components corresponding to two inequivalent spin ladders are plotted.  (c) ESR linewidth (halfwidth at halfheight) vs. magnetic susceptibility log-log plot below 10~K, dashed line corresponds to the cubic dependence as described in the text.  \label{fig:dimpy-width}}
\end{figure}

Quasi-one-dimensional magnet  \dimpy{} (also abbreviated as DIMPY) is an example of a spin-ladder magnet. Crystal structure of this compound is shown at Fig.~\ref{fig:dimpy-struct}. Pairs of chains of copper ions form ladder type structures running along the $a$-axis of the monoclinic crystal. DIMPY structure includes two types of spin ladders, which are bound by second order screw axis. Each of the ladder structures includes an inversion center in the middle of the rung. This inversion ensures that the $g$-tensors for all ions within the given ladder are exactly the same and forbids Dzyaloshinskii-Moriya interaction on the rungs of the ladder. Dzyaloshinskii-Moriya interaction is allowed on the legs of the ladder, it is uniform along the ladder and Dzyaloshinskii-Moriya vectors are exactly opposite on the legs of the given ladder.

\dimpy{} was proved to be almost ideal realization of the spin ladder both in the series of magnetization measurements \cite{dimpy-magn1,dimpy-magn2,dimpy-magn3} and in the inelastic neutron scattering experiments \cite{dimpy-ins,dimpy-ins1,dimpy-ins2}. The energy gap equals $\Delta=0.33$~meV, first critical field is about  30~kOe and the saturation field  $H_{c2}\simeq 300$~kOe. Comparison of the measured neutron scattering spectra with the DMRG modeling \cite{dimpy-ins2,schmidiger-phd} proved that DIMPY is a rare realization of a strong-leg spin-ladder. Exchange integral along the leg is equal to 1.42~meV and that along the rung of the spin ladder equals 0.82~meV \cite{dimpy-ins2}. Field induced ordering was observed in  \dimpy{} at low temperatures, highest ordering temperature in the field of approx. 150~kOe was about 300~mK \cite{dimpy-magn2}.

ESR in DIMPY was studied in details in \cite{glazkov-dimpy}. Samples were grown from solution by a method of slow diffusion in a temperature gradient. As grown samples have a developed plane normal to the $b$-axis and were elongated along the $a$-axis.  Under applied magnetic field ladder structures of different types can become inequivalent being differently oriented with respect to the magnetic field. This would result in different fields of resonance absorption for these ladders. In particular, resonance fields for two types of the ladders coincide for $\vect{H}||b$ and $\vect{H}\perp b$, and it was found experimentally that the largest difference of  $g$-factor values for inequivalent ladders was observed for the field aligned along the bisector between the $a$ and $b$ axes. The later direction is labeled as $\vect{H}||N$ at the Figures.

Examples of resonance absorption spectra for  \dimpy{} at various temperatures are shown at Fig.~\ref{fig:dimpy-spectr}. As temperature decreases below approx. 5~K observed ESR absorption  freezes out, indicating presence of an energy gap. As was described above, at  $\vect{H}||b$ we observe single absorption line, while at $\vect{H}||N$ there are two absorption lines of close intensity corresponding to two ladders of different types. At temperature about 1~K fine structure of triplon ESR spectra is resolved. We did not observed singlet-triplet transitions in DIMPY and the field induced antiferromagnetic phase is beyond the reach of our experimental setup.

Temperature dependencies of magnetic susceptibility and ESR intensity  are shown at Fig.~\ref{fig:dimpy-width}-a. There is a broad maximum at the temperatures around 10~K, this temperature is close to the scale of exchange couplings in DIMPY as expected. Activation energy determined from these data at low temperatures was found to be  2.7~K, this value is slightly less than the zero-field gap $\Delta=0.33$~meV since the magnetic field applied in ESR experiment slightly reduces  energy of the lower triplet sublevel.

ESR linewidth (halfwidth at halfheight) measured in DIMPY over the whole temperature range from 0.45~K to 300~K is shown at Fig.~\ref{fig:dimpy-width}-b. Above 100~K ESR linewidth increases on heating due to activation of spin-phonon relaxation processes. Below 100~K ESR line broadens on cooling, this broadening can be phenomenologically described by the law $\Delta H=\Delta H_0 \left(1+\Theta/T\right)$ with characteristic temperature $\Theta=15...20$~K \cite{glazkov-dimpy}, which is close to the exchange coupling constants values. Below 10~K magnetic susceptibility starts to freeze out, ESR linewidth also decreases on cooling. This decrease of ESR linewidth indicates that spin relaxation at this temperature range is due to triplon-triplon interaction. Peak of linewidth observed at approx. 1~K in one of the field orientations is an artefact of the data processing due to unresolved fine structure of the absorption spectrum. Broadening of the ESR line observed at the lowest temperature below 1~K probably is due to the closeness to the critical point (we recall here that field-induced ordering is observed in DIMPY at very low temperatures \cite{dimpy-magn2}).

To analyze triplon-triplon interactions in more details we plot ESR linewidth vs. ESR intensity (or magnetic susceptibility) in the temperature range from 1~K to 10~K (Fig.~\ref{fig:dimpy-width}-c). Log-log plot clearly demonstrates power law $\Delta H\propto I^3$, pointing to the importance of four-triplon processes for the spin relaxation in DIMPY at low temperatures.

Analysis of the ESR linewidth angular dependencies  established that the main spin relaxation channel in  \dimpy{} is due to uniform Dzyaloshinskii-Moriya interaction, the length of Dzyaloshinskii-Moriya vector is estimated as  $\left|\vect{D}\right|\simeq 0.3$~K \cite{glazkov-dimpy}. Another indication of the importance of Dzyaloshinskii-Moriya interaction for the spin dynamics in DIMPY is  observation of the additional high-frequency ESR mode \cite{ozeroff}.

Thus, it is possible to state a compact problem of the temperature dependence of spin relaxation in a strong-leg spin ladder \dimpy{} below 100~K, where spin-spin interactions are responsible for spin relaxation. Minimal model includes two exchange coupling constants and the length of Dzyaloshinskii-Moriya vector as the parameters. One dimensional models considered earlier  \cite{oshiaff,furuya-jpsj} are not directly applicable to the case of DIMPY and the description of the spin relaxation in a spin ladder with uniform Dzyaloshinskii-Moriya interaction remains an open problem.

\subsection{Effect of nonmagnetic dilution on spin dynamics of the spin ladder magnet DIMPY: creation of $S=1/2$ paramagnetic centers and suppression of the spin-relaxation channel with dilution.}

\begin{figure}
\centering
\includegraphics[width=\figwidth]{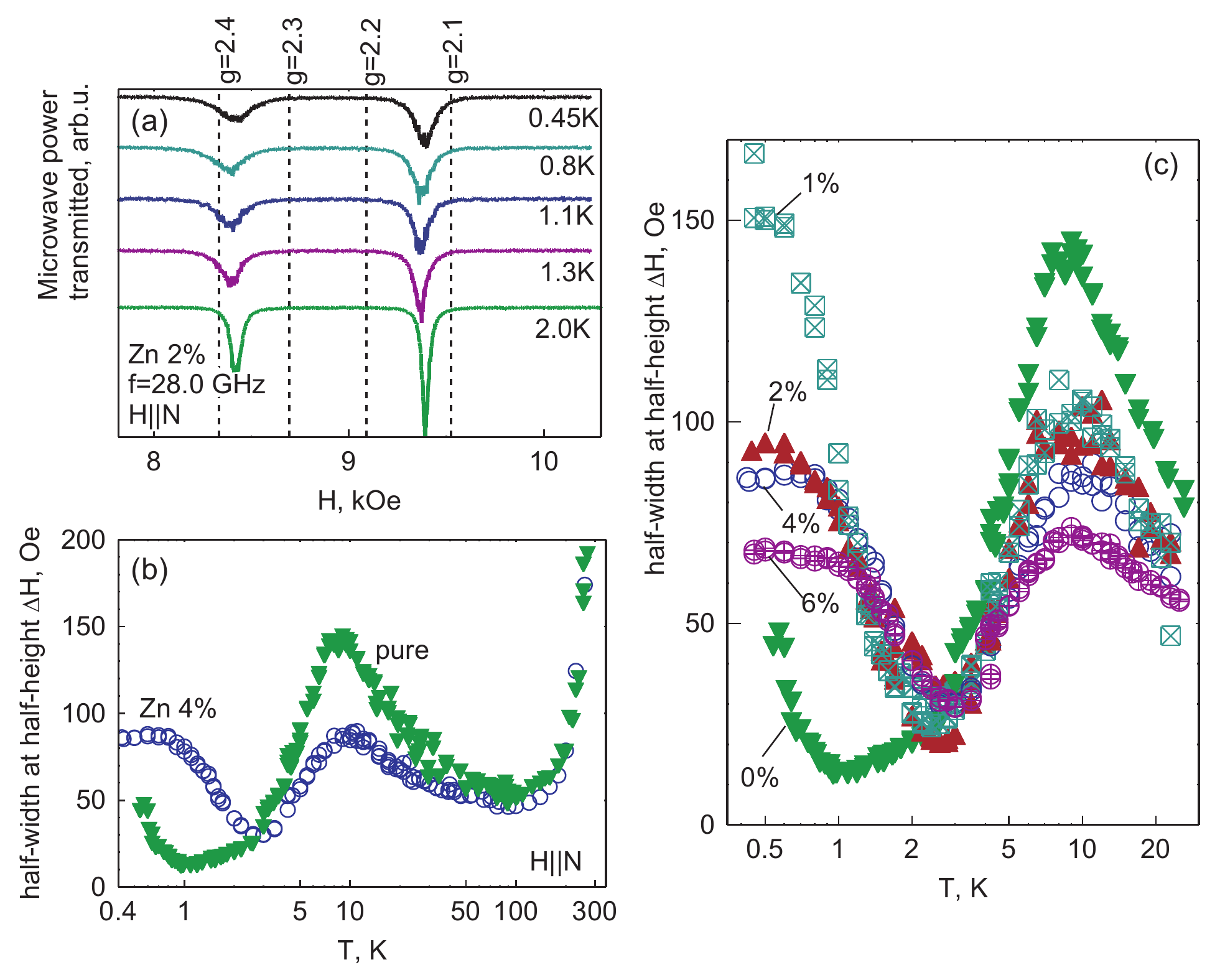}
\caption{(a) Examples of ESR absorption spectra in  \dimpyzn{} with impurity concentration $x=2$\%. $\vect{H}||N$ (b) Comparison of temperature dependencies of ESR linewidth (halfwidth at halfheight) for pure  \dimpy{} and diamagnetically diluted \dimpyzn{} with  4\% Zn content over full temperature range 0.45---300~K. (c) Comparison of the temperature dependencies of ESR linewidth (halfwidth at halfheight) for \dimpyzn{} sample with different Zn content at $T<25$~K.   Data on panels(b) and (c) correspond to right component of ESR absorption at $\vect{H}||N$.      \label{fig:dimpy-doped}}
\end{figure}

It is possible to grow samples of a strong-leg spin-ladder magnet DIMPY with partial substitution of magnetic copper ions by nonmagnetic zinc ions: \dimpyzn{} with $x\leq 6$\% \cite{dimpy-islands}. Effect of diamagnetic dilution is evident within weakly coupled dimers approximation: depletion of spin ladder results in the appearance of free $S=1/2$ spin within the ``broken'' dimer. Inter-dimer coupling ``spreads'' this spin around the defect site on the lengths of the order of magnetic correlation length. This scenario is close to the known case of formation of extended correlated paramagnetic end states in the fragment of dimerised chain or in the fragment of a Haldane chain \cite{chainend1,chainend2,smirglazjetp}. However, spin-ladder geometry of exchange bonds has one key difference from that problem of finite size chain fragments: presence of two legs in a ladder ensures that antiferromagnetic correlations on both sides of defect are correlated.

Static magnetization measurements and inelastic neutron scattering study of \dimpyzn{} samples demonstrate that gap in the excitations spectrum is not suppressed by dilution \cite{dimpy-islands}. In the same time, it was found that the low temperature magnetization does not freeze out completely, indicating presence of paramagnetic centers created by dilution. These paramagnetic centers were found to interact with each other via ``polarization'' of the spin-gap matrix between them. This effective coupling gives rise to additional low-energy dynamics observed below the gap in inelastic neutron scattering experiments \cite{dimpy-islands}. Accounting for this interaction allowed to describe magnetization curves for the full set of \dimpyzn{} samples with $x\leq 6$\% without additional fitting parameters \cite{dimpy-islands}. The later result also proves that nominal and real concentration of zinc in \dimpyzn{} are practically the same.

Effects of diamagnetic dilution on spin dynamics in \dimpyzn{} were studied with ESR spectroscopy in Refs.~\onlinecite{dimpy-doped-esr,dimpy-doped-esr2}. Examples of ESR absorption spectra in diamagnetically diluted DIMPY at different temperatures at $\vect{H}||N$ (at this field direction absorption signals from two types of inequivalently oriented ladders are well resolved) are shown at Fig.~\ref{fig:dimpy-doped}-a. Contrary to the case of pure DIMPY (Fig.~\ref{fig:dimpy-spectr}) samples with $x\neq 0$\% demonstrate intensive ESR absorption at the base temperature of 0.45~K. This allows to identify observed low temperature absorption as ESR from dilution-created paramagnetic centers. Position of low temperature resonance absorption coincides with high-temperature resonance field, which proves that these paramagnetic centers are formed by the copper ions of the unperturbed matrix of the parent compound. Absence of the fine structure proves that these paramagnetic centers have spin $S=1/2$. Intensity of the observed low-temperature absorption can be scaled with magnetic susceptibility \cite{dimpy-islands}, which allows to estimate concentration of paramagnetic centers. This concentration turns out to be comparable with nominal zinc content and the discrepancies can be accounted by taking into account effective interaction between these centers \cite{dimpy-doped-esr}.

Comparison of ESR linewidth (halfwidth at halfheight) in pure and diamagnetically diluted samples is shown at Fig.~\ref{fig:dimpy-doped}-b. One can distinguish two regimes with significantly different linewidth behavior: at low temperatures ($T<4$~K) Curie contribution of dilution-created paramagnetic centers dominate and ESR linewidth in diluted compound is higher than that in pure compound, at higher temperatures $T>4$~K, when thermally activated excitations  of the gapped matrix dominate magnetic response of pure and diluted samples, on the contrary, ESR linewidth in diluted samples turn out to be \emph{smaller} that in the pure DIMPY.

This dependence can be followed for different zinc contents (Fig.~\ref{fig:dimpy-doped}-c). One can see that on increasing zinc content effect is regularly manifested. In low-temperature regime (dominated by paramagnetic centers) the  largest ESR linewidth is observed for the sample with minimal non-zero zinc content. In this case dilution-created paramagnetic centers are well separated by nonmagnetic spin-gap matrix and spin relaxation is determined by some interactions close to the defect, which can be very different from relaxation mechanisms in pure DIMPY. On increasing zinc content low-temperature linewidth decreases through exchange narrowing mechanism due to effective coupling of paramagnetic centers. In high-temperature regime behavior of ESR linewidth in diluted DIMPY is opposite to that in diluted PHCC (Fig.~\ref{fig:phcc-doped}): in the case of \dimpyzn{} introduction of nonmagnetic impurity decreases linewidth, i.e. increases spin relaxation time.

Measurement of angular dependencies of ESR linewidth in \dimpyzn{} demonstrate that these dependencies scale on each other and on similar dependencies for pure DIMPY for all zinc contents studied at the temperatures from 8 to 100~K \cite{dimpy-doped-esr}. This means \cite{oshiaff}, that microscopic mechanism of spin relaxation in high-temperature regime is the same for pure and diamagnetically diluted DIMPY, which is (as it was proved for pure compound \cite{glazkov-dimpy}) the uniform Dzyaloshinskii-Moriya interaction. Thus, effect of diamagnetic dilution on spin dynamics of DIMPY can be interpreted as the suppression of relaxation channel related to the Dzyaloshinskii-Moriya interaction in a certain vicinity of the defect \cite{dimpy-doped-esr}.

\section{Conclusions}

In this review we have considered several examples of application of ESR-spectroscopy technique to the study of collective paramagnets with gapped excitations spectrum (spin-gap magnets). Details of fine structure of triplet sublevels were fully deciphered in a quasi-two dimensional spin-gap magnet \phcc{}, antiferromagnetic resonance absorption in a field-induced ordered phase was observed and the relation of the effective anisotropy of triplet excitations at low fields with anisotropy of the field-induced ordered phase was discussed. On partial substitution of halogen ions mediating the superexchange pathes in \phbc{} we observed formation of unusual paramagnetic centers with spin $S=1$, which is an unusual effect of exchange bond randomness. Quasi-one-dimensional ``spin-tube'' magnet \sul{} demonstrates ESR-active singlet-triplet transition to the point away from the bottom of excitations spectrum, which is due to incommensurate location of the spectrum minimum. Antiferromagnetic resonance absorption observed in \sul{} demonstrates frequency-field dependency typical for helical antiferromagnetic order. Quasi-one dimensional spin-ladder magnet \dimpy{} turns out to be a prominent playground to describe spin relaxation over temperature range  from 0.45~K to 300~K, it was demonstrated that the main spin-spin relaxation channel is due to uniform along the ladder Dzyaloshinskii-Moriya interaction. The later relaxation channel is susceptible to diamagnetic dilution, zinc substitution results in the suppression of this relaxation channel resulting in the decrease of ESR linewidth in \dimpyzn{}. Finally, evidences of triplon-triplon spin relaxation at low temperatures were observed in \phcc{} and \dimpy{}, the details of involved triplon-triplon interactions seems to be strongly system-dependent: while spin relaxation in \phcc{} is a two-triplon interaction process,  spin relaxation in \dimpy{} requires interaction of four triplons.

Observed effects are described within general approaches to spin-gap magnets. However, some  problems still remains an open challenge. In particular, this concerns questions of spin-relaxation in spin-gap magnets: the problem of triplon-triplon interaction in low-temperature regime for different microscopic models and the problem of spin relaxation in a spin-ladder with uniform Dzyaloshinskii-Moriya interaction over the full temperature range from $T\ll\Delta$ to $T\gg J$ (here $\Delta$ is an energy gap and $J$ is a characteristic exchange coupling constant).

\acknowledgements

Author thanks Prof.~A.~Zheludev (ETH-Z\"urich) for his active involvement in the studies mentioned in this review, these researches would not be possible without high quality samples grown in the Laboratory of neutron scattering and magnetism (ETH-Z\"urich) while combination of experimental techniques available in  the Laboratory of neutron scattering and magnetism (ETH-Z\"urich) with ESR spectroscopic facilities of P.~Kapitza Institute (Moscow) allowed to study all relevant effects with high accuracy and reliability.

Author thanks his co-authors T.Yankova (Moscow State University), Yu.~Krasnikova (P.~Kapitza Institute, Moscow), G.~Scoblin (P.~Kapitza Institute, Moscow), D.~H\"{u}vonen (ETH-Z\"urich), E.~Wulf (ETH-Z\"urich), S.~M\"uhlbauer (ETH-Z\"urich), D.~Schmidiger (ETH-Z\"urich), K.~Povarov (ETH-Z\"urich), J.~Sichelschmidt (MPI-CPFS, Dresden) for their contributions to the studies mentioned in this review.

Author thanks Prof.~A.~I.~Smirnov (P.~Kapitza Institute, Moscow), Prof.~L.~E.~Svistov (P.~Kapitza Institute, Moscow), Dr.~A.~B.~Drovosekov (P.~Kapitza Institute, Moscow) and Dr.~T.~A.~Soldatov (P.~Kapitza Institute, Moscow) for their friendly hand with experiment running and numerous useful discussion.

Studies used in this review were carried out with the support from Russian Foundation for Basic Research Grants  15-02-05918,  19-02-00194, Russian Science Foundation Grant 17-12-01505, Swiss National Science Foundation, Division 2. New experimental data on ESR in \phcc{} and \sul{} below 1~K  included in this review were obtained with the support  from Russian Science Foundation Grant 17-12-01505. This review was prepared with the support from Russian Foundation for Basic Research Grant 19-02-00194 and Program of the RAS Presidium ``Actual problems of low temperature physics''.

Crystal structures were drawn with  ``Balls and Sticks'' software \cite{balls-sticks}.

\end{document}